\documentclass[a4paper,12pt]{article}
\pdfoutput=1
\usepackage{graphicx,subfigure,amsmath,amssymb,color}
\usepackage{cite}
\usepackage{CJK}

\newlength{\dinwidth}
\newlength{\dinmargin}
\usepackage[colorlinks = true,
            linkcolor = blue,
            urlcolor  = blue,
            citecolor = red,
            anchorcolor = blue]{hyperref}

\usepackage{bookmark}

\def\la{ \langle}
\def\ra{ \rangle}
\def\r{ \gamma}
\def\u{\mu}
\def\v{\nu}

\def \d {{\rm d}}

\allowdisplaybreaks

\begin{document}
\begin{CJK*}{GB}{gbsn}

\title{\bf Decay constants of pseudoscalar and vector mesons with improved holographic wavefunction}
\author{\hspace{-0.6cm}\small Qin Chang~(常钦)$^{1,2}$, Xiao-Nan Li~(李晓楠)$^{1}$, Xin-Qiang Li~(李新强)$^{2}$,
and Fang Su~(苏方)$^{2}$~\footnote{Corresponding author: sufang@itp.ac.cn}
\\
{ $^1$\small Institute of Particle and Nuclear Physics, Henan Normal University, Henan 453007,  China}\\
{ $^2$\small Institute of Particle Physics and Key Laboratory of Quark and Lepton Physics~(MOE),}\\
{     \small Central China Normal University, Wuhan, Hubei 430079,
China}
}
\date{}
\maketitle

\begin{abstract}
\noindent We calculate the decay constants of light and heavy-light pseudoscalar and vector mesons with improved soft-wall holographic wavefuntions, which take into account the effects of both quark masses and dynamical spins. We find that the predicted decay constants, especially for the ratio $f_V/f_P$, based on light-front holographic QCD, can be significantly improved, once the dynamical spin effects are taken into account by introducing the helicity-dependent wavefunctions. We also perform detailed $\chi^2$ analyses for the holographic parameters ({\it i.e.} the mass-scale parameter $\kappa$ and the quark masses), by confronting our predictions with the data for the charged-meson decay constants and the meson spectra. The fitted values for these parameters are generally in agreement with those obtained by fitting to the Regge trajectories. At the same time, most of our results for the decay constants and their ratios agree with the data as well as the predictions based on lattice QCD and QCD sum rule approaches, with only a few exceptions observed.
\end{abstract}


\newpage

\section{Introduction}

Inspired by the correspondence between string theory in anti-de Sitter (AdS) space and conformal field theory (CFT) in physical space-time~\cite{Maldacena:1997re,Gubser:1998bc,Witten:1998qj}, a class of AdS/QCD approaches with two alternative AdS/QCD backgrounds has been successfully developed for describing the phenomenology of hadronic properties~\cite{Polchinski:2001tt,Karch:2006pv}. In this direction, light-front~(LF) holographic QCD exploits an approximate AdS$_5$/QCD duality to obtain a Schr\"odinger-like equation for the transverse wavefunctions~(WFs) of hadrons with massless quarks~(see, for instance, Refs.~\cite{deTeramond:2008ht,Brodsky:2008pg,deTeramond:2010ge,Brodsky:2014yha} for details), and has been successfully used to predict the spectroscopy of hadrons~\cite{deTeramond:2005su,Brodsky:2006uqa,deTeramond:2009xk,deTeramond:2014asa,Dosch:2015nwa,Brodsky:2016yod,Brodsky:2016rvj,Dosch:2016zdv,Nielsen:2018uyn,Branz:2010ub}, the dynamical observables such as the transition form factors and the structure functions~\cite{Brodsky:2007hb,Brodsky:2008pf,Brodsky:2011yv}, and the behavior of the QCD running coupling in the nonperturbative domain~\cite{Brodsky:2010ur,Deur:2016cxb,Deur:2016tte}.
In this approach, the LF dynamics depends only on the boost-invariant variable chosen as either the invariant mass $M_0$ or the invariant impact variable $\zeta$, and the dynamical properties are encoded in the hadronic LF wavefunction (LFWF), which takes the form~\cite{deTeramond:2008ht}
\begin{equation}
\psi(x,\zeta,\varphi)=e^{iL\varphi}X(x)\frac{\phi(\zeta)}{\sqrt{2\pi\zeta}}\,.
\end{equation}
The LF eigenvalue equation, $P_{\mu}P^{\mu}|\psi \ra=M^2|\psi \ra$, can be then reduced to an effective single-variable LF Schr\"odinger equation for $\phi(\zeta)$,
\begin{equation}\label{eq:scheq}
\left[-\frac{{\rm d}^2}{{\rm d}\zeta^2 }-\frac{1-4L^2}{4\zeta^2}+U(\zeta)\right] \phi(\zeta)=M^2\phi(\zeta)\,,
\end{equation}
which is relativistic, frame-independent and analytically tractable~\cite{deTeramond:2008ht}. This equation provides a first-order approximation to the light-front QCD eigenvalue problem for hadrons in the valence Fock-state representation.

The effective potential $U(\zeta)$ in Eq.~\eqref{eq:scheq}, which acts on the valence Fock states of hadrons and enforces confinement at some scale, is holographically related to a unique dilation profile in  anti-de Sitter~(AdS) space~\cite{deTeramond:2010ge,Brodsky:2014yha}. After holographic mapping, one can arrive at a concise form of a colour-confining harmonic oscillator,  $U(\zeta,J)=\lambda^2\zeta^2+2\lambda(J-1)$, in which $\sqrt{\lambda}=\kappa$ is a mass-scale parameter.
Using such a confining potential, one can then obtain the eigenvalues, corresponding to the squares of the hadron masses, by solving the LF Schr\"odinger equation. In Refs.~\cite{deTeramond:2014asa,Dosch:2015nwa,Brodsky:2016yod,Brodsky:2016rvj}, the observed light meson and baryon spectra are successfully described by extending the superconformal quantum mechanics to the light-front and its embedding in AdS space. Moreover, similar analyses are further extended to the heavy-light hadrons~\cite{Dosch:2016zdv,Nielsen:2018uyn}.

The eigensolution of Eq.~\eqref{eq:scheq} gives the holographic WF, which encodes the dynamical properties of the considered hadrons, and is given explicitly as~\cite{Brodsky:2007hb,Brodsky:2008pg}
\begin{equation} \label{eq:LFWFb}
\psi_{n,L}^{(0)}=\frac{1}{N}e^{iL\varphi}\sqrt{x(1-x)}\zeta^LL^L_n(|\lambda|\zeta^2)e^{-|\lambda|\zeta^2/2},
\end{equation}
for a meson with LF angular momentum $L$ and radial excitation number $n$. Here $L^L_n$ is the associated Laguerre polynomials, $N=\sqrt{(n+L)!/(n!\pi)}|\lambda|^{(L+1)/2}$ the normalization, and $\zeta^2=x(1-x){\bf b}_{\bot}^2$ with ${\bf b}_{\bot}$ being the invariant transverse impact variable and $x$ the momentum fraction. This holographic WF has been widely used to evaluate the hadronic observables~\cite{Brodsky:2007hb,Branz:2010ub,Brodsky:2011yv,Hwang:2012xf,Braga:2015jca,Vega:2009zb}.

It should be noted, however, that quark masses are not taken into account in the holographic WF given by Eq.~\eqref{eq:LFWFb}. Keeping quarks  massless is essential for reducing the dynamics to a single-variable problem; this is also required by the underlying conformal symmetry of QCD, and exhibits an exact agreement of the AdS equation of motion with the LF Hamiltonian~\cite{Brodsky:2014yha}. In addition, the helicity indices have also been suppressed in Eq.~\eqref{eq:LFWFb}, which is legitimate if the helicity dependence of the holographic WF decouples from the dynamics. Both the limit of massless quarks and the assumption of helicity independence
are actually consistent with the semi-classical approximation within which the AdS/QCD correspondence is exact~\cite{Brodsky:2014yha,deTeramond:2005su,Brodsky:2006uqa}. For realistic phenomenological applications, however, it is essential to restore both the quark-mass and helicity dependences of the WF, to improve the predictions of holographic QCD compared to data~\cite{Brodsky:2008pg,Vega:2009zb,Swarnkar:2015osa,Ahmady:2016ufq,Chang:2016ouf,Ahmady:2016ujw}.  In this paper, taking these two effects into account, we will revisit  the decay constants of pseudoscalar and vector mesons in the holographic QCD framework.

The decay constant is an important ingredient in applying QCD to hard exclusive processes via the factorization theorem~\cite{Lepage:1980fj,Efremov:1979qk,Chernyak:1983ej}, and provides essential information on the QCD interaction between the valence quark and anti-quark of the involved mesons. It also provides a direct source of information on the Cabibbo-Kobayashi-Maskawa matrix elements~\cite{Koppenburg:2017mad} and plays a significant role in the neutral-meson mixing processes~\cite{Artuso:2015swg}. In LF holographic QCD, assuming that the helicity dependence of the holographic WF decouples from the dynamics, one can derive a simple factorizable formula for the decay constant~\cite{Brodsky:2007hb,Vega:2009zb},
\begin{equation} \label{eq:nlfDC}
f_P=f_V=2\sqrt{2N_c}\int_0^1\d x\int\frac{\d^2{\bf k}_\bot}{16\pi^3}\,\psi(x,\mathbf{k}_{\bot})\,,
\end{equation}
for both pseudoscalar and vector mesons. In Eq.~\eqref{eq:nlfDC}, the holographic WF in the $\mathbf{k}_{\bot}$ space, $\psi(x,\mathbf{k}_{\bot})$, is obtained via Fourier transform from  Eq.~\eqref{eq:LFWFb}, with the explicit form given by~\cite{Brodsky:2007hb,Brodsky:2014yha}
\begin{eqnarray}\label{eq:LFWFkT}
\psi(x,\mathbf{k}_{\bot})=\,\frac{4\pi}{\kappa}\frac{1}{ \sqrt{x(1-x)}}\,e^{-\frac{\mathbf{k}_{\bot}^2}{2\kappa^2\,x(1-x)}}\,.
\end{eqnarray}
It is obvious from Eqs.~\eqref{eq:nlfDC} and \eqref{eq:LFWFkT} that valid results for different $(q\bar{q}^{\prime})$ bound states are possibly predicted only when the quark-mass correction to the holographic WF $\psi(x,\mathbf{k}_{\bot})$ is considered.

A simple generalization of Eq.~\eqref{eq:LFWFkT} for massive quarks follows from the Brodsky-T\'{e}ramond ansatz~\cite{Brodsky:2008pg}, which assumes that the momentum-space holographic WF is a function of the invariant off-energy-shell quantity, rather than only of the transverse momentum. This leads to the following replacement in Eq.~\eqref{eq:LFWFkT}:
\begin{eqnarray}\label{eq:replac}
K_0\equiv\frac{\mathbf{k}_{\bot}^2}{\,x(1-x)}\,\to\,K=K_0+m_{12}^2\,,\qquad m_{12}^2\equiv\frac{m_1^2}{x}+\frac{m_2^2}{1-x}\,,
\end{eqnarray}
where $m_{1}$ and $m_{2}$ are the masses of quark and anti-quark in a $(q_1\bar{q}_2)$ Fock state, respectively. It has been demonstrated that predictions based on the modified holographic WF for massive quarks improve the description of data on the electromagnetic and photon-to-meson transition form factors for $\pi$ and $\eta^{(\prime)}$ mesons~\cite{Swarnkar:2015osa}. However, for the heavy-light mesons, it has been found that the magnitude of the decay constants is grossly underestimated with increasing heavy-quark mass, because in this case the longitudinal momentum fraction carried by the light quark is pushed to a very small value~\cite{Dosch:2016zdv}. In order to remedy this evidently too strong suppression, the heavy-quark mass term is further modified through the replacement $m_Q^2\to\alpha^2m_Q^2$ with $\alpha\sim0.5$ ($Q$ denotes the relatively heavier quark in a two-quark bound state)~\cite{Dosch:2016zdv}. More generally, as suggested in Ref.~\cite{Teramond:GHP2009}, the quark-mass term in the exponential of the holographic WF can be absorbed into the longitudinal mode, $f(x,m_1,m_2)$; as analysed in Refs.~\cite{Branz:2010ub,Lyubovitskij:2010zz,Vega:2011ck,Gutsche:2014oua}, the mass-scale parameter $m_{12}$ entering in $f(x,m_1,m_2)$ should not necessarily be identified with the parameter $\kappa$ characterizing the dilation field. In this paper, this kind of generalized holographic WF for massive quarks will be further studied and confronted with the decay constants of light and heavy-light mesons.

However, even with the quark mass included in the modified WF, the holographic QCD prediction based on Eqs.~\eqref{eq:LFWFb}--\eqref{eq:replac} still results in an unsatisfactory relation $f_V/f_P=1$ for a given $(q_1\bar{q}_2)$ pseudoscalar (P) and vector (V) multiplicity. This is obviously disfavored by the experimental data, for instance~\cite{Rosner:2015wva,PDG,Straub:2015ica,Grossmann:2015lea},
\begin{equation}
 f_{\rho}/f_{\pi}\approx1.615>1\,,\quad  f_{K^*}/f_{K}\approx1.308>1\,.
\end{equation}
Such a tension indicates further possible improvement of the holographic WF, such as by taking into account the dynamical spin effects. In Refs.~\cite{Forshaw:2011yj,Forshaw:2012im,Ahmady:2012dy,Ahmady:2013cva,Ahmady:2013cga,Ahmady:2014sva,Ahmady:2014cpa,Ahmady:2015gva,Ahmady:2015yea,Ahmady:2016ufq,Chang:2016ouf,Ahmady:2016ujw}, the dynamical ({\it i.e.} momentum-dependent) spin WF, $S_{h,\bar{h}}$ has been introduced to restore the helicity dependence of the holographic WF. Even though such an improvement seems to be phenomenologically successful, there exists arbitrariness more or less in determining the explicit form of $S_{h,\bar{h}}$. For a particular hadronic state with assigned $J^{PC}$ quantum numbers, the spin-orbit wavefunction can be obtained uniquely by the interaction-independent Melosh transformation~\cite{Melosh:1974cu} from the ordinary equal-time static one, if we treat $\psi(x,\mathbf{k}_{\bot})$ as the radial wavefunction in the LF space~\cite{Chung:1988my,Jaus:1989au}. In this paper, we will account for the dynamical spin effects in detail by using the helicity-dependent WF obtained through the Melosh transformation, and show that the holographic QCD predictions for $f_P$, $f_V$ and $f_V/f_P$ can be significantly improved.

Our paper is organized as follows. In Section 2, the theoretical framework and calculation for the decay constants with the modified holographic WF are presented. Our numerical results and discussions are then given in Section 3. Finally, we give our conclusions in Section 4.

\section{Theoretical framework}

\subsection{Improvements of holographic WF}

As suggested in Ref.~\cite{Teramond:GHP2009}, a general form of the soft-wall holographic WF including the quark-mass term for a given $(q_1\bar{q}_2)$ ground state can be written as~\cite{Branz:2010ub}
\begin{eqnarray}\label{eq:modLFWF}
\psi(x,\mathbf{k}_{\bot})=\,\frac{4\pi}{\kappa}\frac{1}{ \sqrt{x(1-x)}}\,e^{-\frac{\mathbf{k}_{\bot}^2}{2\kappa^2\,x(1-x)}}\,f(x,m_1,m_2)\,,
\end{eqnarray}
with the longitudinal mode given by~\cite{Branz:2010ub}
\begin{eqnarray}\label{eq:longitudinalmode}
f(x,m_1,m_2)\equiv N f(x) e^{-\frac{m_{12}^2}{2\lambda_{12}^2}}\,,
\end{eqnarray}
where $f(x)=1$ in our case, $\lambda_{12}$ is a mass-scale parameter, and $N$ is the normalization constant determined by the condition
\begin{eqnarray}\label{eq:NFact}
\int_0^1\d x f^2(x,m_1,m_2)=1\,.
\end{eqnarray}
The holographic WF, $\psi(x,\mathbf{k}_{\bot})$, given by Eq.~\eqref{eq:modLFWF}, with $N$ fixed by Eq.~\eqref{eq:NFact}, automatically satisfies  the usual normalization condition
\begin{eqnarray}\label{eq:norpsi}
 \int_0^1 \d x  \int \frac {\d^2\mathbf{k}_{\bot}}{2(2\pi)^3}|\psi(x,\mathbf{k}_{\bot})|^2=1\,.
\end{eqnarray}

The dimensional parameter $\lambda_{12}$ introduced in Eq.~\eqref{eq:longitudinalmode} should not necessarily be identified with the dilation parameter $\kappa$\cite{Branz:2010ub,Lyubovitskij:2010zz,Vega:2011ck,Gutsche:2014oua}. While the simplification $\lambda_{12}=\kappa$ is generally allowed for light hadrons, $\lambda_{12}>\kappa$ is required to remedy the strong suppression caused by the heavy-quark mass $m_Q$~\cite{Dosch:2016zdv}. Fitting to the mass spectra and the decay constants of heavy-light mesons, the authors of Ref.~\cite{Branz:2010ub} have shown that $\lambda_{12}$ scales generally as ${\cal O}(m^{1/2}_{Q})$ with a universal value of the dilation parameter $\kappa=0.55~{\rm GeV}$. It is therefore expected that $\lambda_{12}\to\kappa$ in the limit of $m_Q\to0$, and becomes significantly large with increasing $m_Q$. In addition, as found in Refs.~\cite{Branz:2010ub,Vega:2011ck,Gutsche:2014oua}, a relatively larger value for $\kappa$ is also required in order to fit better the  heavy-light hadron spectra. Based on the above observations, we will choose in this paper $\lambda_{12}/\kappa_{12}=1.1\,,1.7\,,3.5\,,3.5$ for  $Q=q\,,s\,,c\,,b$ for simplicity, in which $q= u,d$ and a subscript ``$12$" is added to the parameter $\kappa$ to clarify its difference for different $(q_1\bar{q}_2)$ mesonic states.

As the description of the motion of the constituents in terms of the inner momentum vectors is independent of the motion of the system as a whole, the wavefunction of a bound state must be a simultaneous eigenfunction of the mass operator as well as the angular momentum operators $\mathbf{J}^2$ and $\mathbf{J}_3$, and should depend, therefore, only on the inner momentum vectors and spins of the constituents. This implies that the wavefunction for a $(q_1\bar{q}_2)$ bound state with a given spin $J$ should be spin dependent. As a result, the helicity-dependent LFWF in $\mathbf{k}_{\bot}$ space can be generally written as
\begin{eqnarray}\label{eq:LFWFP2}
\Psi_{h,\bar{h}}(x,\mathbf{k}_{\bot})=S_{h,\bar{h}}(x,\mathbf{k}_{\bot}) \psi(x,\mathbf{k}_{\bot}) \,,
\end{eqnarray}
where $\psi(z,\mathbf{k}_{\bot})$ has already been given by Eq.~\eqref{eq:modLFWF}, while $S_{h,\bar{h}}(x,\mathbf{k}_{\bot})$, with $h$~($\bar{h}$) being the helicity of the (anti-)quark, is the helicity-dependent wavefunction obtained by the interaction-independent Melosh transformation~\cite{Melosh:1974cu} from the ordinary equal-time static one, and constructs a state with definite $(J,J_3)$ out of the light-front helicity eigenstates $(h,\bar{h})$.

Explicitly, the covariant form of the spin-orbit wavefunction can be written as~\cite{Jaus:1989au,Choi:1997iq,Choi:2007yu,Hwang:2010hw}
\begin{eqnarray}\label{eq:wfS}
S_{h,\bar{h}}(x,\mathbf{k}_{\bot})=\frac{\bar{u}(k_1,h)\,(\not\!\bar{P}+M_0)\times\Gamma\, v(k_2,\bar{h})}{\sqrt{2}\bar{M}_0(M_0+m_1+m_2)}\,,
\end{eqnarray}
where $\bar{P}\equiv k_1+k_2$, and $\bar{M}_0^2\equiv M_0^2-(m_1-m_2)^2$, with $M_0^2=\frac{m_1^2+\mathbf{k}_{\bot}^2}{x_1} +\frac{m_2^2+\mathbf{k}_{\bot}^2}{x_2}$ being the invariant mass squared of a $(q_1\bar{q}_2)$ bound state. For the pseudoscalar and vector mesons, we have
\begin{eqnarray}\label{eq:gama}
\Gamma_P=\gamma_5,\qquad \Gamma_V=-\not\!\tilde{\epsilon}\,,
\end{eqnarray}
with the polarization vectors given explicitly by
\begin{eqnarray}\label{eq:polvc}
\tilde{\epsilon}_0&=&\frac{1}{M_0}\left(P^+,\frac{-M_0^2+{\bf P}_{\bot}^2}{P^+},{\bf P}_{\bot}\right)\,,\\
\tilde{\epsilon}_{\pm}&=&\left( 0,\frac{2}{P^+}\boldsymbol{\epsilon}_{\bot}\cdot {\bf P}_{\bot}, \boldsymbol{ \epsilon}_{\bot}\right)\,, \quad\boldsymbol{ \epsilon}_{\bot}\equiv \mp\frac{1}{\sqrt{2}}(1,\pm i)\,.
\end{eqnarray}
While the transverse polarization vectors  $\tilde{\epsilon}_{\pm}$ coincide with $\epsilon_{\pm}$ of the vector meson, the longitudinal one $\tilde{\epsilon}_0$ is different from $\epsilon_0$ of the vector meson, with the latter given by
\begin{eqnarray}\label{eq:polvcM}
\epsilon_0=\frac{1}{M}\left(P^+,\frac{-M^2+{\bf P}_{\bot}}{P^+},{\bf P}_{\bot}\right)\,,
\end{eqnarray}
where $M$ and $P^\mu=\left(P^+,P^-,{\bf P}_{\bot}\right)$ are the physical mass and four-momentum of the meson, respectively.
Here, we would like to emphasize that the covariant helicity-dependent wavefunction $S_{h,\bar{h}}(x,\mathbf{k}_{\bot})$ given by Eq.~\eqref{eq:wfS} can automatically satisfy the normalization condition
\begin{eqnarray}\label{eq:SNor}
\sum_{h\,\bar{h}}S_{h,\bar{h}}^{\dagger}(x,\mathbf{k}_{\bot}) S_{h,\bar{h}}(x,\mathbf{k}_{\bot})=1\,.
\end{eqnarray}
This in turn indicates that the total normalization condition
 \begin{eqnarray}\label{eq:nortot}
\sum_{h,\bar{h}}\int_0^1 \d x \int\frac{\d^2\mathbf{k}_{\bot}}{2(2\pi)^3} |\Psi_{h,\bar{h}}(x,\mathbf{k}_{\bot})|^2=1\,,
\end{eqnarray}
is also automatically satisfied by the helicity-dependent LFWF  $\Psi_{h,\bar{h}}(x,\mathbf{k}_{\bot})$ defined by Eq.~\eqref{eq:LFWFP2}.

In analogy with the leading-order helicity structure of the photon WF~\cite{Lepage:1980fj}, the authors of Refs.~\cite{Forshaw:2011yj,Forshaw:2012im,Ahmady:2012dy,Ahmady:2013cva,Ahmady:2013cga,Ahmady:2014sva,Ahmady:2014cpa,Ahmady:2015gva,Ahmady:2015yea} assume a simple form of the helicity WF for the vector meson, $S_{h,\bar{h}}=N \frac{\bar{u}(k_1,h)}{x}\not\!\epsilon\, \frac{v(k_2,\bar{h})}{1-x}$, with the dimensional constant $N$ determined by Eq.~\eqref{eq:nortot}. Obviously, in such an analogy, the assumed form of $S_{h,\bar{h}}$ for the vector meson is incomplete because the photon is massless. 

\subsection{Decay constants with improved holographic WF}

The decay constants are defined by
\begin{eqnarray}
\label{eq:dcp}
\la 0 | \bar q_2\r^\u\r_5 q_1|P(p)\ra=if_P p^\u \,,
\end{eqnarray}
for a pseudoscalar meson, and
\begin{eqnarray}
\label{eq:dcv}
&\la 0 | \bar q_2\r^\u q_1|V(p,\lambda)\ra=
f_VM_V \epsilon^\u_{\lambda} \,,\\
\label{eq:dcvT}
&\la 0 | \bar q_2\sigma^{\u\v} q_1|V(p,\lambda)\ra = if_V^T( \epsilon^\u_{\lambda} P^\v- \epsilon^\v_{\lambda} P^\u)\,,
\end{eqnarray}
for a vector meson with longitudinal ($\lambda=0$) and transverse ($\lambda=\pm$) polarizations, respectively.

In the framework of LF quantization, adopting the Lepage-Brodsky (LB) conventions and the light-front gauge~\cite{Brodsky:1997de,Lepage:1980fj}, and working in the leading valence Fock-state approximation, we can expand a mesonic eigenstate $|M\ra$ by the noninteracting two-particle Fock states as
\begin{equation}
|M\ra = \sum_{h,\bar{h}} \int \frac{\d k^+ \d^2{\bf k_{\bot}}}{(2\pi)^32\sqrt{k^+(P^+-k^+)}} \Psi_{h,\bar{h}}\left(k^+/P^+,{\bf k}_{\bot}\right)|k^+,k_{\bot},h;P^+-k^+,-k_{\bot},\bar{h}\ra \,.
\label{eq:Fockexp}
\end{equation}
With the LF helicity spinors $u_h$ and $v_{h}$, the dynamical Dirac (quark) field is expanded as~\cite{Brodsky:2014yha,Brodsky:1997de}
\begin{equation}
\psi_+(x)= \int \frac{\d k^+}{\sqrt{2k^+}}\frac{ \d^2{\bf k}_{\bot}}{(2\pi)^3} \sum_h \left[b_{h} (k) u_h(k)e^{-ik\cdot x} + d^{\dagger}_{h} (k) v_{h} (k)e^{ik\cdot x}\right] \,,
\label{eq:qfexp}
\end{equation}
in terms of particle creation and annihilation operators, which satisfy the equal LF-time anti-commutation relations
\begin{equation}
\{b^{\dagger}_{h} (k), b_{h'} (k^{\prime}) \}=\{d^{\dagger}_{h} (k), d_{h'} (k^{\prime}) \}= (2\pi)^3  \delta(k^+-k'^{+})\delta^2({\bf k}_{\bot}-{\bf k}'_{\bot}) \delta_{h h'}.
\label{anticommutation}
\end{equation}

Using the above formulae, we can generally express the left-hand-side of Eqs.~\eqref{eq:dcp}--\eqref{eq:dcvT} as
\begin{eqnarray}
 \la 0 | \bar{q} \Gamma^{\prime} q|M\ra =
\sqrt{N_c}\, \sum_{h,\bar{h}}\, \int \frac{\d x\d^2{\bf k}_{\bot}}
{(2\pi)^32\sqrt{x\bar{x}}}\,
\psi(x,{\bf k}_{\bot})\times S_{h,\bar{h}}(x,{\bf k}_{\bot})\,
\bar{v}_{\bar{h}}(\bar{x},-{\bf k}_{\bot}) \Gamma^{\prime}  u_h(x,{\bf k}_{\bot})\,,
\label{eq:meder1}
\end{eqnarray}
where $\bar{x}=1-x$, and $\Gamma^{\prime}=\r^\u\r_5$, $\r^\u$ and $\sigma^{\u\v}$, corresponding respectively to Eqs.~\eqref{eq:dcp}, \eqref{eq:dcv} and \eqref{eq:dcvT}. Taking the plus component ($\mu=+$) of the currents from Eqs.~\eqref{eq:dcp} and \eqref{eq:dcv}, and plugging in the spin-orbit wavefunction given by Eq.~\eqref{eq:wfS}, we finally arrive at
\begin{eqnarray}
\label{eq:redcp}
f_P&=&\frac{\sqrt{N_c}}{\pi}\int_0^1\d x\int\frac{\d^2{\bf k}_{\bot}}{(2\pi)^2}\frac{ \psi(x,{\bf k}_{\bot})}{\sqrt{x\bar{x}}}\frac{1}{\sqrt{2}\bar{M}_0}
\left(\bar{x}m_1+xm_{2}\right) \,,\\[0.2cm]
\label{eq:redcv}
f_V&=&\frac{\sqrt{N_c}}{\pi}\int_0^1\d x\int\frac{\d^2{\bf k}_{\bot}}{(2\pi)^2}\frac{ \psi(x,{\bf k}_{\bot})}{\sqrt{x\bar{x}}}\frac{1}{\sqrt{2}\bar{M}_0}
\left(\bar{x}m_1+xm_{2}+\frac{2{\bf k}_{\bot}^2}{M_0+m_1+m_2}\right) \,,
\end{eqnarray}
in which we take the $\lambda=0$ component for evaluating $f_V$.
For $f_V^T$, taking $\mu=+$, $\lambda=\pm$ and multiplying both sides of Eq.~\eqref{eq:dcvT} by $\epsilon^*_{\v}$, we can obtain
\begin{eqnarray}
\label{eq:redcvT}
f_V^T&=&\frac{\sqrt{N_c}}{\pi}\int_0^1\d x\int\frac{\d^2{\bf k}_{\bot}}{(2\pi)^2}\frac{ \psi(x,{\bf k}_{\bot})}{\sqrt{x\bar{x}}}\frac{1}{\sqrt{2}\bar{M}_0}
\left(\bar{x}m_1+xm_{2}+\frac{{\bf k}_{\bot}^2}{M_0+m_1+m_2}\right) \,.
\end{eqnarray}
While the decay constants $f_P$ and $f_V$ can be extracted from experiment through the decays $P^-\to \ell^- \bar{\nu}_{\ell}(\gamma)$, $V^0\to \ell^+\ell^-$ and $\tau^-\to M^-\nu_{\tau}$~\cite{Rosner:2015wva,PDG,Straub:2015ica,Grossmann:2015lea,Neubert:1997uc,Ball:2006eu}, the transverse one $f_V^T$ is not that easily accessible in experiment and hence has to be estimated theoretically. It is also noted that $f_V^T$ is scale dependent due to the nonzero anomalous dimension of the tensor current. In the holographic QCD framework, the scale dependence of $f_V^T$ can be roughly identified by introducing an ultraviolet cut-off on the transverse momenta, {\it i.e.}, $\int\d^2{\bf k}_{\bot}\to \int^{|{\bf k}_{\bot}|<\mu}\d^2{\bf k}_{\bot}$~\cite{Kogut:1973ub,Diehl:2002he}. As has been found in, for instance, Refs.~\cite{Forshaw:2011yj,Forshaw:2012im,Ahmady:2012dy,Ahmady:2013cva,Ahmady:2013cga,Ahmady:2014sva,Ahmady:2014cpa}, the scale evolution of $f_V^T$ is not significant when $\mu>1~{\rm GeV}$; therefore, the predictions based on Eq.~\eqref{eq:redcvT} should be viewed to hold only at some low-energy scale $\mu\sim1~{\rm GeV}$. Results at higher scales can be obtained from $f_V^T(1~{\rm GeV})$ through the leading-order renormalisation-group improved relation
\begin{eqnarray}
f_V^T(\mu)=f_V^T(1{\rm GeV}) \left[ \frac{\alpha_s(\mu)}{\alpha_s(1{\rm GeV})}\right]^{\frac{C_F}{\beta_0}}\,,
\end{eqnarray}
where $C_F=(N_C^2-1)/(2N_C)$ and $\beta_0=11-2/3n_f$, with $N_C$ and $n_f$ being the number of colours and flavours, respectively.

From the theoretical expressions, Eqs.~\eqref{eq:redcp}, \eqref{eq:redcv} and \eqref{eq:redcvT}, for the decay constants of mesons composed of the same ($q_1\bar{q}_2$) constituents, we can make the following qualitative observations:
\begin{itemize}
\item Comparing Eqs.~\eqref{eq:redcp} and \eqref{eq:redcv}, our results for $f_P$ and $f_V$ indicate the experimentally favored relation $f_V>f_P$, which is significantly different from the traditional result $f_V=f_P$ implied by Eq.~\eqref{eq:nlfDC}. Our predictions for the decay constants can, therefore, be improved once the dynamical spin effect is taken into account in the LFWF.

\item In the heavy quark limit, the dynamical spin effect becomes trivial and the traditional result given by Eq.~\eqref{eq:nlfDC} is, therefore, expected to be recovered from Eqs.~\eqref{eq:redcp}--\eqref{eq:redcvT}. This can be inferred from the following analyses. Assuming $q_1$ to be a heavy quark, which implies that $m_1\gg m_2$ and $m_1^2 \gg {\bf k}_{\bot}^2$, we can then neglect safely the terms proportional to ${\bf k}_{\bot}^2$ in the bracket of  Eqs.~\eqref{eq:redcv} and \eqref{eq:redcvT}, leading to the same residual $(\bar{x}m_1+xm_2)$ (or $\bar{x}m_1$ if $m_2$ is also neglected) in the numerator of Eqs.~\eqref{eq:redcp}--\eqref{eq:redcvT}. With the same approximation, on the other hand, one can easily find that the denominator can be simplified as
    \begin{eqnarray}
    \sqrt{x\bar{x}}\bar{M}_0=\sqrt{(\bar{x}m_1+xm_2)^2+{\bf k}_{\bot}^2}\simeq (\bar{x}m_1+xm_2)\,,
    \end{eqnarray}
    which cancels exactly the residual in the bracket of  Eqs.~\eqref{eq:redcp}--\eqref{eq:redcvT}. Therefore, one can finally find that our results given by Eqs.~\eqref{eq:redcp}, \eqref{eq:redcv} and \eqref{eq:redcvT} all coincide with the traditional result given by Eq.~\eqref{eq:nlfDC}, and $f_P=f_V=f_V^T$ in the heavy quark limit, which is generally expected in the heavy quark effective theory~\cite{Neubert:1993mb}.

\item From Eqs.~\eqref{eq:redcp}, \eqref{eq:redcv} and \eqref{eq:redcvT}, one can also find an interesting relation
    \begin{eqnarray}
    f_P+f_V=2f_V^T\,,
    \end{eqnarray}
    which is a consequence of the LF approach~\cite{Choi:2007yu,Hwang:2010hw}. Such a relation agrees surprisingly well with the old ${\rm SU}(6)$ symmetry relation~\cite{Leutwyler:1973mu}. Moreover, it is also generally followed in the lattice QCD~(LQCD) and QCD sum rules (QCDSR) approaches.
\end{itemize}
Equipped with the formulae and analyses given above, we will then present our numerical results and discussions in the next section.

\section{Numerical results and discussion}

\subsection{Fit for the holographic parameters}

As is well known, the decay constant, once measured experimentally, would provide a severe test for the adequacy of the wavefunction. In order to determine the values of the holographic parameters, the mass-scale parameter $\kappa$ and the quark masses, we will first perform a detailed $\chi^2$-analysis for these parameters, by confronting our results with the experimentally well-measured charged-meson decay constants $f_{P}$ and $f_{V}$, which are collected in the second column of Table~\ref{tab:dc}. Here $f_P$ and $f_V$ are extracted from the purely leptonic decays $P^{\pm}\to \ell^{\pm} \nu_{\ell}$~\cite{Rosner:2015wva,PDG} and from the one-prong hadronic $\tau$ decays $\tau^{\pm}\to V^{\pm} \nu_{\tau}$~\cite{Straub:2015ica,Grossmann:2015lea}, respectively.

\begin{table}[t]
\begin{center}
\caption{\label{tab:dc} \small Experimental data~\cite{PDG,Straub:2015ica} and theoretical results in LQCD~\cite{Aoki:2016frl,Lubicz:2016bbi,Braun:2016wnx}, QCDSR~\cite{Ball:2006eu,Narison:2015nxh,Lucha:2017sdx,Wang:2015mxa,Gelhausen:2013wia},  LFQM~\cite{Choi:2007yu,Hwang:2010hw}, and this work for the decay constants (in unit of ${\rm MeV}$). The values in bold in the last two columns denote our predictions for the corresponding decay constants, the experimental data (if they exist) of which are not used as constraints during the fits. See text for further details.}
\vspace{0.2cm}
\let\oldarraystretch=\arraystretch
\renewcommand*{\arraystretch}{1.1}
\setlength{\tabcolsep}{8.8pt}
\begin{tabular}{lcccccccccccc}
\hline\hline
                 & data              &LQCD              & QCDSR
                 & LFQM              & SI               & SII\\  \hline
$f_{\pi}$&$130.3\pm0.3$&$130.2 \pm 1.7$&---                    &$131$                  &$130.2^{+3.0}_{-2.8}$  &$130.3^{+3.6}_{-3.3}$ \\
$f_{\rho}$ &$210\pm4$       &$199\pm4$       &$206\pm7$        &$215$                  &$\bf{166^{+2}_{-4}}$   &$210^{+6}_{-6}$ \\
$f_{K}$&$156.1\pm0.5$ & $155.6\pm0.4$ &---                     &$155$                   &$153.4^{+2.8}_{-2.0}$  &$156.4^{+4.4}_{-9.1}$\\
$f_{K^*}$ &$204\pm7$      & ---                      &$222\pm8$       &$223$                  &$\bf{186^{+2}_{-3}}$  &$204^{+7}_{-9}$\\
$f_{D}$ &$203.7\pm4.7$    & $211.9\pm1.1$  &$204.0\pm4.6$ &$206.0\pm8.9$   &$206.5^{+4.9}_{-8.2}$ &$203.5^{+4.6}_{-4.6}$\\
$f_{D^*}$ &---                    & $223.5\pm8.4$   &$250\pm8$      &$259.6\pm14.6$  &$\bf{226.6^{+\phantom{0}5.9}_{-10.2}}$&$\bf{230.1^{+6.2}_{-6.2}}$\\
$f_{D_s}$ &$257.8\pm4.1$ &$249.0\pm1.2$  &$243.2\pm4.9$  &$267.4\pm17.9$ &$233.1^{+5.0}_{-5.4}$&$257.8^{+7.3}_{-5.5}$   \\
$f_{D_s^*}$&---                 &$268.8\pm6.6$    &$290\pm11$      &$338.7\pm29.7$ &$\bf{254.7^{+6.3}_{-6.7}}$&$\bf{289.7^{+6.3}_{-4.5}}$   \\
$f_{B}$  &$188\pm25$      &$187.1\pm4.2$    &$204.0\pm5.1$   &$204\pm31$      &$193.4^{+\phantom{0}4.7}_{-10.6}$   &$187.2^{+4.0}_{-4.3}$ \\
$f_{B^*}$ &---                   &$185.9\pm7.2$     &$210\pm6$        &$225\pm38$       &$\bf{198.7^{+\phantom{0}4.9}_{-11.3}}$   &$\bf{193.1^{+4.3}_{-4.6}}$ \\
$f_{B_s}$ &---                   &$227.2\pm3.4$    &$234.5\pm4.4$    &$281\pm54$      &$225.5^{+6.2}_{-7.2}$ &$227.1^{+6.6}_{-5.2}$\\
$f_{B_s^*}$&---                 &$223.1\pm 5.4$   &$221\pm 7$        &$313\pm67$      &$\bf{231.9^{+6.6}_{-7.6}}$ &$\bf{234.0^{+6.4}_{-5.2}}$ \\
\hline\hline
\end{tabular}
\end{center}
\end{table}

It is also known that the measured meson masses can put another strong constraint on the holographic parameters~\cite{deTeramond:2005su,Brodsky:2006uqa,deTeramond:2009xk,
deTeramond:2014asa,Dosch:2016zdv,Branz:2010ub,Lyubovitskij:2010zz,Vega:2011ck,Gutsche:2014oua}. To this end, we will adopt the following master formula for the meson masses~\cite{Branz:2010ub,Lyubovitskij:2010zz}
\begin{eqnarray}\label{eq:MnJ}
M_{nJ}^2=4\kappa^2(n+\frac{L+J}{2})+\int_0^1\d x\, m_{12}^2 \,f^2(x,m_1,m_2)+\Delta M_C^2\,,
\end{eqnarray}
where the first term reflects the limit of parity doubling between vector and axial mesons, and the second is due to the inclusion of the longitudinal mode $f(x,m_1,m_2)$, which accounts for the quark-mass dependence of the soft-wall holographic WF. The last term in Eq.~\eqref{eq:MnJ} results from the contribution of an additional colour Coulomb-like potential due to the one-gluon exchanges between quarks~\cite{Sergeenko:1994ck,Gershtein:2006ng}, and reads~\cite{Branz:2010ub,Lyubovitskij:2010zz}
\begin{eqnarray}
\Delta M_C^2=-\frac{64\alpha_s^2(\mu_{12}^2) m_1m_2}{9(n+L+1)^2}\,,
\end{eqnarray}
with $\mu_{12}=2m_1m_2/(m_1+m_2)$. The strong coupling $\alpha_s(\mu_{12}^2)$ depends on the number of quark flavours involved, $N_f$, and takes the ``freezing" form~\cite{Badalian:2004xv,Ebert:2009ub}
\begin{eqnarray}
\alpha_s(\mu^2)=\frac{12\pi}{(33-2N_f)\ln\frac{\u^2+M_B^2}{\Lambda^2}}\,,
\end{eqnarray}
where $\Lambda$ is the QCD scale parameter, and $M_B$ the background mass. Numerically, we take as input $\Lambda=(420\pm 5)~{\rm MeV}$ and $M_B=(855\pm10)~{\rm MeV}$~\cite{Branz:2010ub,Lyubovitskij:2010zz}. The shift of $M^2$ due to $\Delta M_C^2$ is negative and proportional to the quark mass squared. This implies that the term $\Delta M_C^2$ plays an important role in constraining the holographic parameters.

\begin{figure}[t]
\begin{center}
\subfigure[]{\includegraphics[scale=0.5]{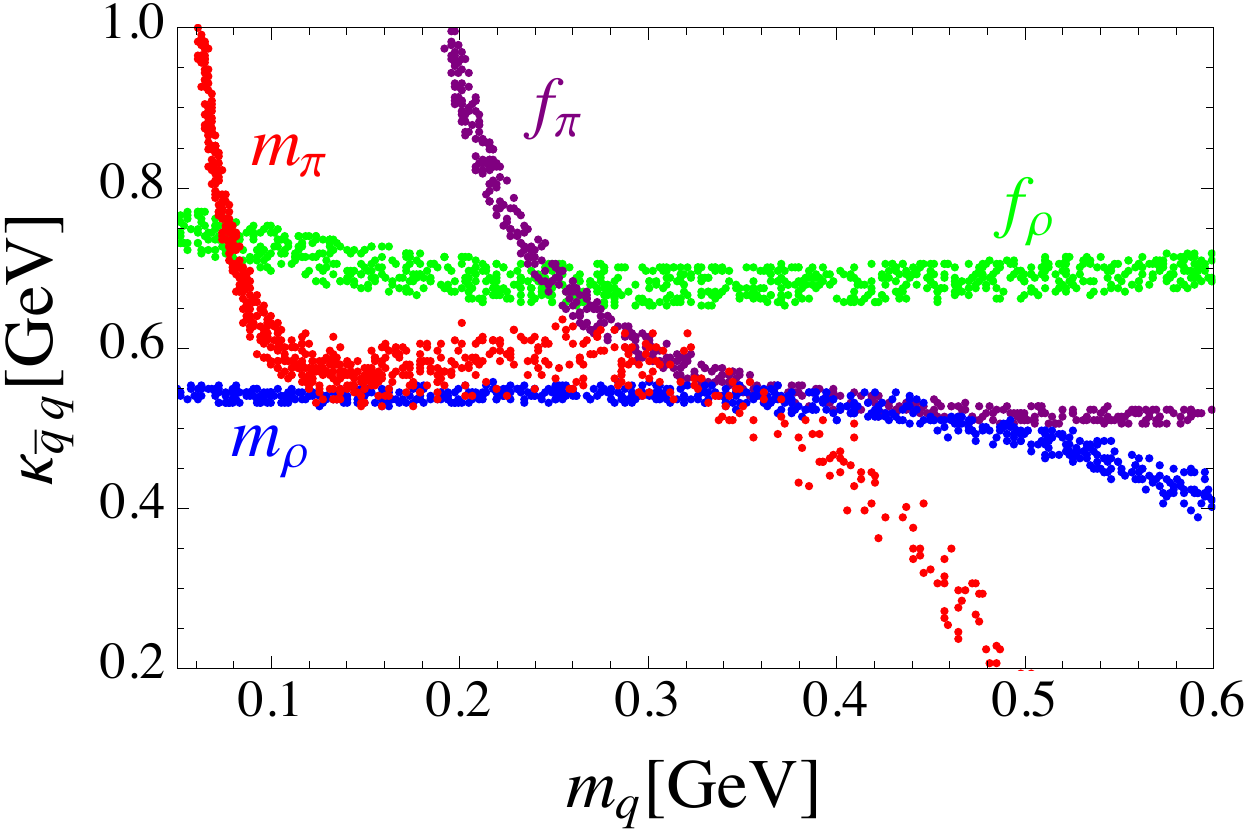}}\qquad
\subfigure[]{\includegraphics[scale=0.5]{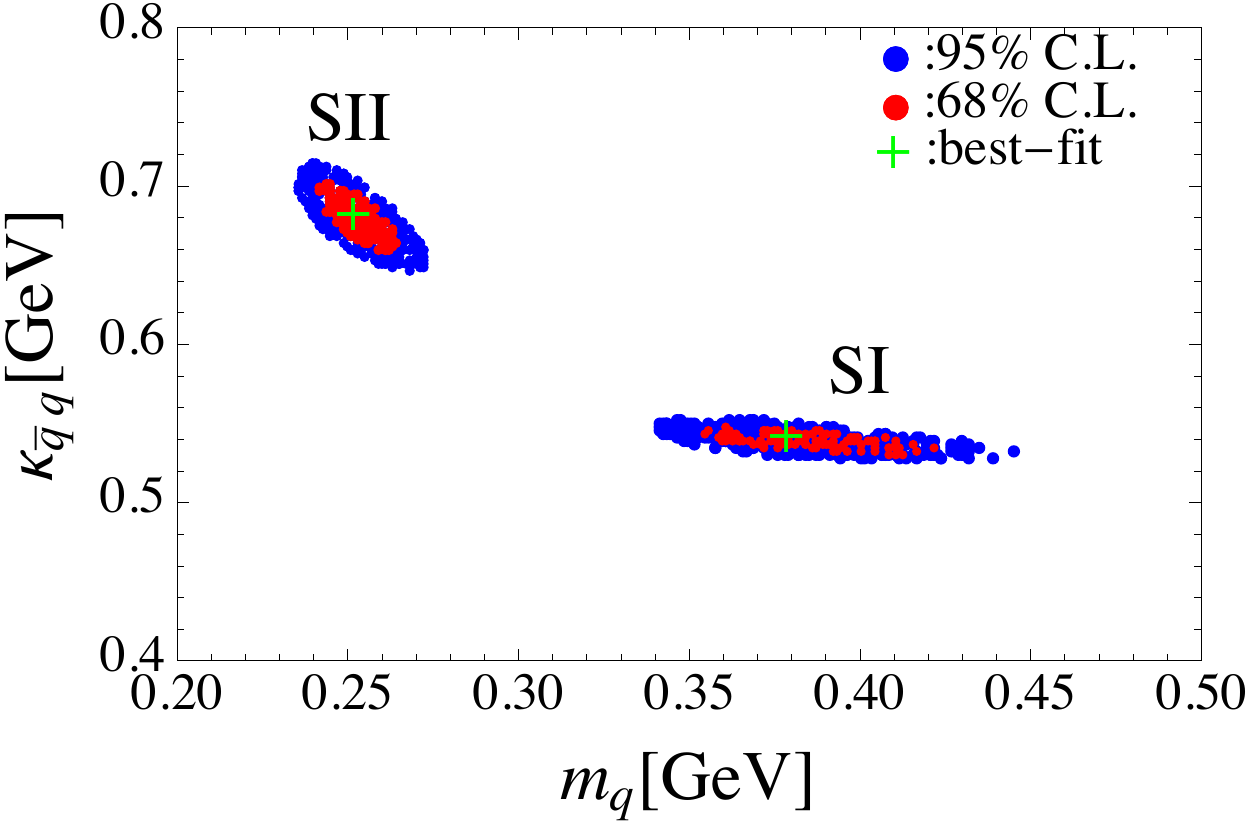}}
\caption{\label{fig:fitud} \small The fitted spaces for $\kappa_{\bar{q}q}$ and $m_q$ under the separate constraints from $f_{\pi,\rho}$ and $m_{\pi,\rho}$ at $95\%$ C.L. (a), as well as under their combined constraint (b). See text for further explanation.}
\end{center}
\end{figure}

Under the separate constraints from the decay constants $f_{\pi,\rho}$ and the masses $m_{\pi,\rho}$, the allowed spaces for $\kappa_{\bar{q}q}$ and $m_q$, with $q=u,d$, are shown in Fig.~\ref{fig:fitud}(a). We can see that a small $\kappa_{\bar{q}q}\sim 0.54~{\rm GeV}$ with $m_q\sim0.33~{\rm GeV}$ is favored by $m_{\pi,\rho}$ and $f_{\pi}$, but disfavored by $f_{\rho}$; however, a relatively large $\kappa_{\bar{q}q}\sim0.68~{\rm GeV}$ with $m_q\sim0.25~{\rm GeV}$ is favored by $f_{\pi,\rho}$, but disfavored especially by $m_{\rho}$. Such a tension is caused mainly by the different requirements for the dilation parameter $\kappa_{\bar{q}q}$ from $f_{\rho}$ and $m_{\rho}$. On the other hand, even though our result for $f_\rho$ is still lower than the experimental data, $f_{\rho}^{\rm exp.}=210~{\rm GeV}$~\cite{Straub:2015ica}, the observed tension has been significantly moderated compared to the case obtained without including the dynamical spin effect, $f_{\rho}=170~{\rm GeV}$~\cite{Lyubovitskij:2010zz}. Motivated by these observations, we will divide our detailed fits and analyses into two scenarios dubbed scenario~I~(SI) and scenario~II (SII), respectively, together with the following two comments:
\begin{itemize}
\item For the decay constant $f_{V}$, there has been a debate about the zero-mode~\cite{Burkardt:1992sz,Brodsky:1998hn,deMelo:1998an,Choi:1998nf}  contribution to the matrix element of the weak current defined by Eq.~\eqref{eq:dcv}, in the standard LF~(SLF) formalism. In Refs.~\cite{Jaus:1999zv,Jaus:2002sv}, Jaus claimed that the zero-mode contribution to $f_{V}$ cannot be avoided even for the case of the plus component of the weak current; while the authors of Refs.~\cite{Bakker:2002mt,Bakker:2003up,Choi:2013oxa} found that this contribution may be model dependent, especially on the form of the meson vertex operator. Our result, Eq.~\eqref{eq:redcv}, is obtained in the zero-binding-energy limit ({\it i.e.} the four-momenta of the meson and its constituents are all on-mass-shell); while the manifestly covariant LF~(CLF) approach allows a nonzero binding energy and leads to the result (the zero-mode contribution is now included)~\cite{Jaus:1999zv,Cheng:2003sm,Choi:2013mda}
    \begin{eqnarray} \label{eq:codcv}
    f_V^{\rm CLF}&=&\frac{\sqrt{N_c}}{4\pi^3}\int_0^1\d x\int\d^2{\bf k}_{\bot}\frac{ \psi(x,{\bf k}_{\bot})}{\sqrt{x\bar{x}}}\frac{1}{\sqrt{2}\bar{M}_0}
    \nonumber\\
    &&\times\frac{1}{M}\left[x\bar{x}M^2+{\bf k}_{\bot}^2+m_1m_2+x(m_1+m_{2})\frac{{\bf k}_{\bot}^2+m_2^2-\bar{x}^2M^2}{\bar{x}(M_0+m_1+m_2)}\right] \,
    \end{eqnarray}
    obtained with the plus component of the current and the longitudinal polarization vector $\epsilon^\mu_{0}$. Numerically, we find a $\sim20\%$ enhancement for $f_V$ obtained with Eq.~\eqref{eq:codcv} compared to that with Eq.~\eqref{eq:redcv}. However, the enhancement reduces to be $\sim10\%$ when using the results obtained with the perpendicular components of the current and the transverse polarization vector $\epsilon^\mu_{\pm}$~\cite{Cheng:2003sm}. Although such an enhancement is favored by the current data, the CLF approach is plagued by the self-consistency problem, {\it i.e.}, $f_{V\,,\lambda=0}^{\rm CLF}\neq f_{V\,,\lambda=\pm}^{\rm CLF}$~\cite{Cheng:2003sm}. The same problem also exists in the SLF formalism followed in this paper,  {\it i.e.}, $f_{V\,,\lambda=0}^{\rm SLF}\neq f_{V\,,\lambda=\pm}^{\rm SLF}$, with the latter obtained by taking the combination $(\mu=\bot, \lambda=\pm)$. Interestingly, such a problem can be ``resolved'' by using the ``Type II" correspondence proposed in Ref.~\cite{Choi:2013mda}, and it is found numerically that $f_{V\,,\lambda=0}^{\rm CLF}= f_{V\,,\lambda=\pm}^{\rm CLF}= f_V^{\rm SLF}$~\cite{Choi:2013mda}. It is also noted that our result given by Eq.~(\ref{eq:redcv}) does not change under the ``Type II" replacement. If so, our result for $f_V$ is acceptable. However, the physical origin of the ``Type II" correspondence is still unclear, even though it is helpful to maintain the self-consistency of the LF formalisms.

Therefore, in SI, we consider only the constraints from $f_{P}$ and $m_{P,\,V}$, but leave $f_{V}$ as our predictions, due to the above issues complicated by the zero-mode contribution and the self-consistency problem.

\item In deriving the master formula for the meson masses, Eq.~\eqref{eq:MnJ}, we include only the one-gluon exchange contribution to the effective potential $U(\zeta)$. However, further corrections to $U(\zeta)$ exist. For instance, the hyperfine-splitting potential~\cite{Zhou:2003gj,Karliner:2006fr,Karliner:2008sv} was found to provide addition small negative (positive) contributions to $M^2$ for pseudoscalar (vector) mesons~\cite{Branz:2010ub,Lyubovitskij:2010zz}. In Ref.~\cite{Grigoryan:2010pj}, an additional constant term was added to the effective potential, to control the masses of the ground state. In fact, we find that any modification to the effective potential may significantly affect the light-meson masses. For example, if the hyperfine-splitting contribution is included, the result $\kappa_{\bar{q}q}\sim0.68~{\rm GeV}$ with $m_q\sim0.25~{\rm GeV}$  will be allowed by $m_{\pi}$ even though it is still disfavored by $m_{\rho}$.

    Therefore, due to the above issues of possible further modifications to the meson masses, and in order to show clearly the dependence of the decay constant on the holographic parameters, we consider in SII only the constraints from $f_{P}$ and $f_V$, while discarding those from $M^2$.
\end{itemize}

Our final $\chi^2$-fitting results for the holographic parameters in both SI and SII are shown in Figs.~\ref{fig:fitud}(b), \ref{fig:qs}, \ref{fig:qc} and \ref{fig:qb}~\footnote{In our $\chi^2$-fits, as a conservative choice, an additional $1\%$ error is assigned to the experimental data if its significance is larger than $100~\sigma$. In addition, the LQCD results for $f_{B_d}$ and $f_{B_s}$ are used in the fits due to the lack of  corresponding experimental data.}, with the corresponding best-fit results summarized in Table~\ref{tab:hp}, in which the results for $\kappa$ obtained by fitting to the Regge trajectories~\cite{Brodsky:2014yha,Dosch:2016zdv} are also listed for comparison. With the fitted holographic parameters given in Table~\ref{tab:hp}, our theoretical results for the decay constants are then collected in Table~\ref{tab:dc}, in which the predictions based on LQCD~\cite{Aoki:2016frl,Lubicz:2016bbi,Braun:2016wnx}, QCDSR~\cite{Ball:2006eu,Narison:2015nxh,Lucha:2017sdx,Wang:2015mxa,Gelhausen:2013wia} and  LFQM~\cite{Choi:2007yu,Hwang:2010hw} approaches are also listed for comparison. The following two subsections are devoted to our detailed analyses for a given $(q_1\bar{q}_2)$ state.

\begin{table}[t]
\begin{center}
\caption{\label{tab:hp} \small Fitted results for the parameter $\kappa$ and quark masses (in unit of ${\rm GeV}$) in both SI and SII. The results for $\kappa$ obtained by fitting to the Regge trajectories~\cite{Brodsky:2014yha,Dosch:2016zdv} are also given for comparison.}
\vspace{0.2cm}
\renewcommand*{\arraystretch}{1.1}
\setlength{\tabcolsep}{5pt}
\begin{tabular}{lccccccccc}
\hline\hline
       &$\kappa_{\bar{q}q}$           & $\kappa_{\bar{q}s}$ & $\kappa_{\bar{q}c}$ & $\kappa_{\bar{s}c}$  & $\kappa_{\bar{q}b}$& $\kappa_{\bar{s}b}$\\ \hline
SI& $0.540^{+0.007}_{-0.010}$ &$0.602^{+0.007}_{-0.006}$  &$0.765^{+0.032}_{-0.018}$ & $0.836^{+0.020}_{-0.021}$& $0.918^{+0.014}_{-0.034}$& $0.994^{+0.020}_{-0.022}$\\
SII&$0.680^{+0.021}_{-0.021}$ &$0.674^{+0.026}_{-0.020}$  &$0.783^{+0.020}_{-0.020}$ & $0.942^{+0.018}_{-0.012}$& $0.892^{+0.013}_{-0.014}$&$0.975^{+0.011}_{-0.011}$ \\
Refs.~\cite{Brodsky:2014yha,Dosch:2016zdv}& $[0.54,0.59]$ &$[0.54,0.59]$ &$[0.655,0.736]$&$[0.735,0.766]$&$[0.963,1.13]$&$[1.11,1.16]$\\
\hline\hline
       &$m_q$                                 & $m_s$                                 &$m_c$ & $m_b$\\ \hline
SI& $0.379^{+0.042}_{-0.024}$&$0.594^{+0.007}_{-0.027}$ &$1.64^{+0.05}_{-0.03}$ & $5.17^{+0.10}_{-0.03}$ \\
SII&$0.252^{+0.012}_{-0.010}$&$0.593^{+0.158}_{-0.101}$  &$1.5$ &$4.8$ \\
\hline\hline
\end{tabular}
\end{center}
\end{table}

\subsection{Light mesons}

Under the combined constraint from $\pi$ and $\rho$ mesons, the fitted results in both SI and SII are shown in Fig.~\ref{fig:fitud}(b). For the mass-scale parameter $\kappa_{\bar{q}q}$~($q=u,d$), it is found that our result in SI, $\kappa_{\bar{q}q}^{\rm SI}=0.540^{+0.007}_{-0.010}\,{\rm GeV}$, agrees remarkably well with the result $[0.54,0.59]\,{\rm GeV}$ obtained by fitting to the Regge trajectories of $(\bar{q}q)$ states~\cite{Brodsky:2014yha}. However, compared with these results, a relatively larger value, $\kappa_{\bar{q}q}^{\rm SII}=0.680^{+0.021}_{-0.021}\,{\rm GeV}$, in SII is required to fit $f_{\rho}$. This implies that SII might be refuted unless there exists an unknown negative potential for vector mesons. For the light-quark mass, on the other hand, it is found that our result, $m_q^{\rm SI}\sim 0.379 \,{\rm GeV}$, is much larger than that obtained in Ref.~\cite{Brodsky:2014yha}, due to the inclusion of the negative colour Coulomb-like potential contribution; such a relatively large light-quark mass is also favored by the decay constants of light pseudoscalar mesons, which can be seen from Eq.~\eqref{eq:redcp}.

Using the best-fit values of $\kappa_{\bar{q}q}$ and $m_q$ obtained in SI and SII, we get $1.28\,{\rm (SI)}$ and $1.61\,{\rm (SII)}$ for the ratio $f_{\rho}/f_{\pi}$. Although still smaller than the data $1.62$~\cite{PDG,Straub:2015ica}, our result $1.28\,{\rm(SI)}$ has in fact  been significantly improved compared with the traditional one, $f_{\rho}/f_{\pi}=1$, obtained without considering the dynamical spin effect. Furthermore, using Eq.~\eqref{eq:codcv} to include the zero-mode contribution, we obtain $f_{\rho}/f_{\pi}=1.58\,{\rm (SI)}$, agreeing well with the data. This implies that the zero-mode contribution is possibly important and worth further theoretical investigation. In addition, our predictions for the ratio $f_{\rho}^T({\rm 2\,GeV})/f_{\rho}$, $0.78\,{\rm (SI)}$ and $0.71\,{\rm (SII)}$~\footnote{Here the NLO scaling factor $f_{V}^T({\rm 2\,GeV})/f_{V}^T({\rm 1\,GeV})=0.876$ has been used.}, are also comparable with those obtained in the LQCD and QCDSR approaches, for instance, $f_{\rho}^T({\rm 2\,GeV})/f_{\rho}=0.76$~\cite{Jansen:2009hr}, $0.63$~\cite{Braun:2016wnx}~(LQCD), and $f_{\rho}^T({\rm 2\,GeV})/f_{\rho}=0.72$~\cite{Ball:1996tb,Ball:1998kk,Ball:2004rg}, $0.69\pm0.04$~\cite{Ball:2006nr}~(QCDSR).

\begin{figure}[t]
\begin{center}
\subfigure[]{\includegraphics[scale=0.5]{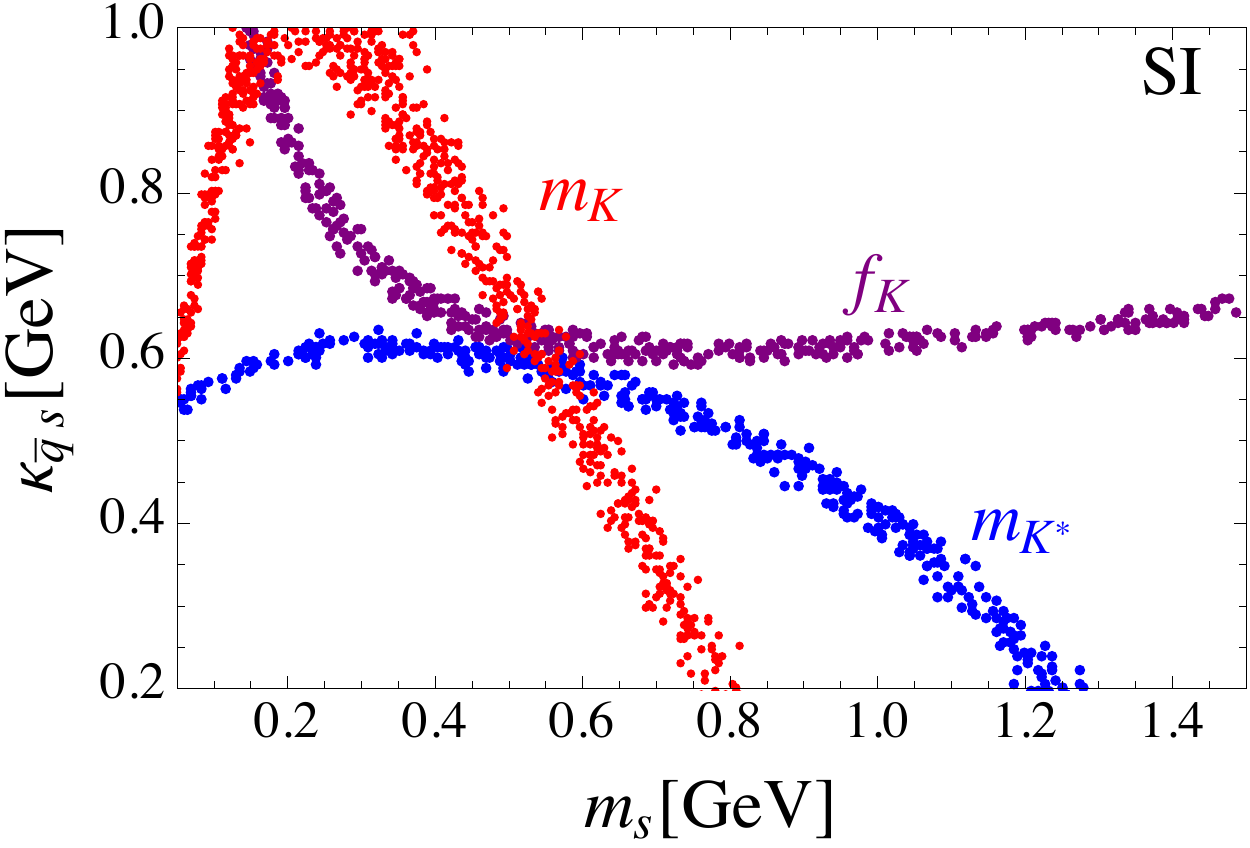}}\qquad
\subfigure[]{\includegraphics[scale=0.5]{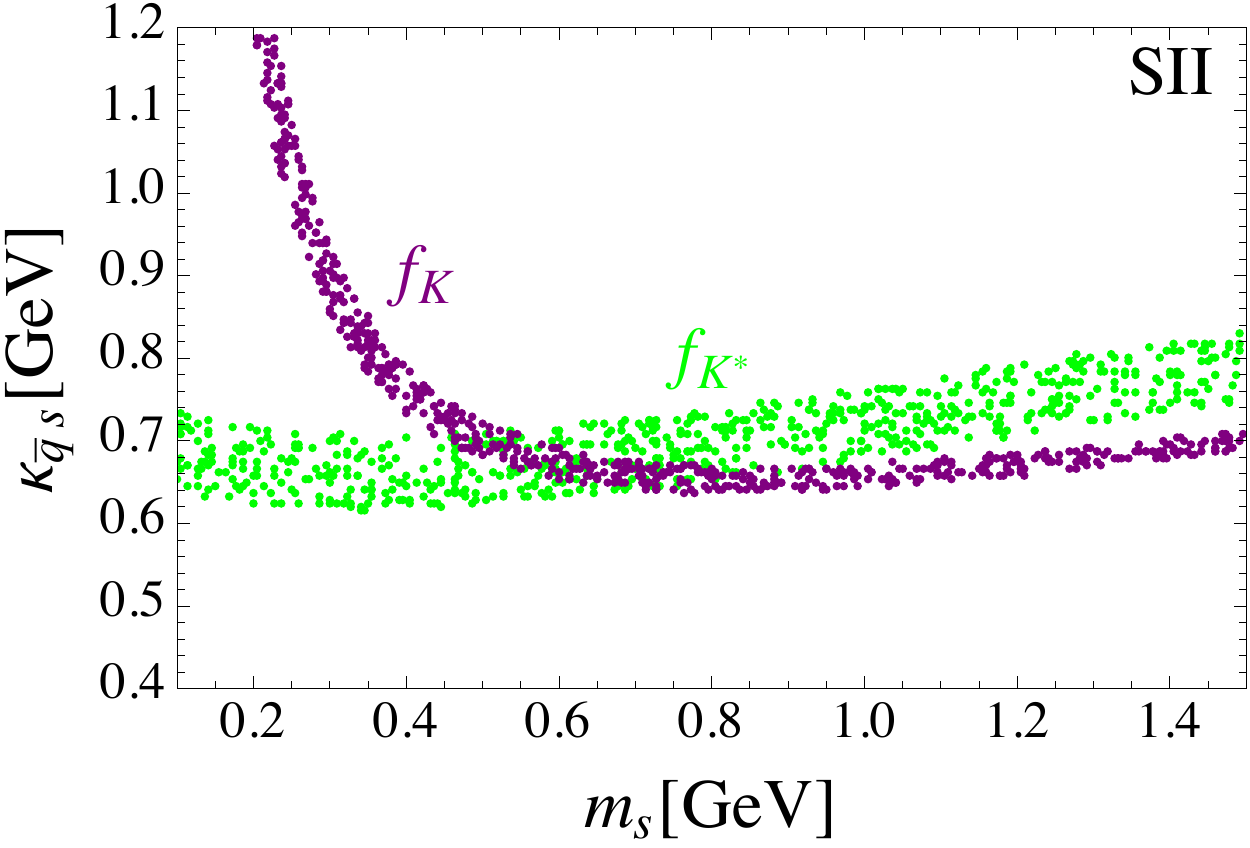}}\\
\subfigure[]{\includegraphics[scale=0.5]{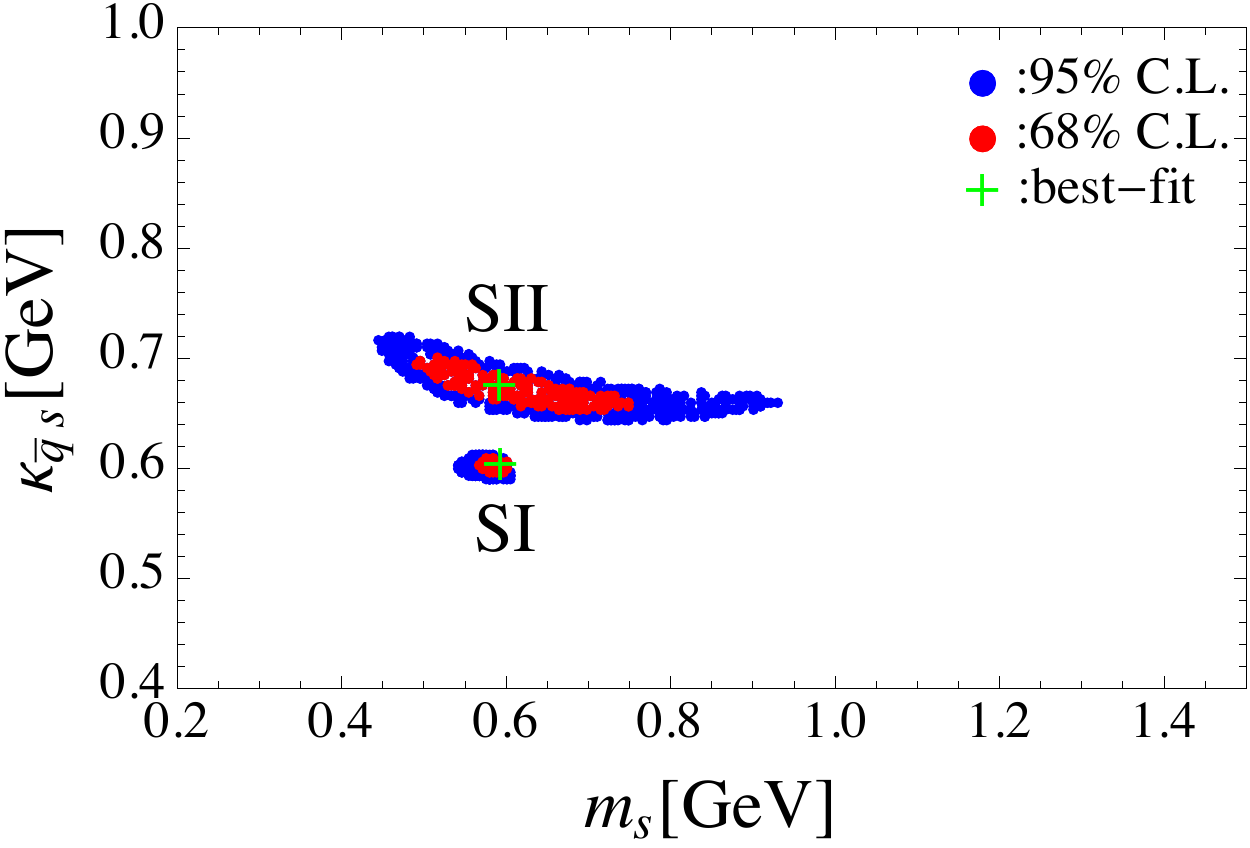}}
\caption{\label{fig:qs} \small The fitted spaces for $\kappa_{\bar{q}s}$  and $m_s$ under the separate constraints from the masses and decay constants of $K$ and $K^*$ mesons at $95\%$ C.L. in SI~(a) and SII~(b), as well as under their combined constraint~(c). See text for further explanation. }
\end{center}
\end{figure}

With the best-fit values of $m_q$ obtained in SI and SII as inputs, the fitted results for the $(s\bar{q})$ system are shown in Fig.~\ref{fig:qs}. From Fig.~\ref{fig:qs}(a), it can be seen clearly that the parameters in SI are strictly bounded at $\kappa_{\bar{q}s}\sim0.6~{\rm GeV}$ with $m_s\sim 0.59~{\rm GeV}$ under the constraints from $f_{K}$, $m_{K}$ and $m_{K^*}$. At the same time, compared with SI, a relatively large $\kappa_{\bar{q}s}\sim0.67~{\rm GeV}$ is required by $f_{K^*}$ in SII, as shown by Fig.~\ref{fig:qs}(b). Such a situation is similar to what has been observed in the $\pi$ and $\rho$ systems, but the tension between SI and SII for $\kappa_{\bar{q}s}$ is not so serious. The final combined fitting results are shown in Fig.~\ref{fig:qs}(c). Numerically, we find  that our fitted result $\kappa_{\bar{q}s}= 0.602\,{\rm GeV}\,(\rm SI)$ is consistent with the result $[0.54,0.59]\,{\rm GeV}$ obtained by fitting to the Regge trajectories~\cite{Brodsky:2014yha}; it is, however, larger than $\kappa_{\bar{q}q}= 0.540\,{\rm GeV}\,(\rm SI)$. This is explained by the significant flavour-symmetry-breaking effect indicated by the data $f_K/f_{\pi}=1.2$~\cite{PDG}.

It is also found from Table~\ref{tab:dc} that the tension between the theoretical prediction in SI and the data for $f_{K^*}$ is not as serious as that observed for $f_{\rho}$. Using the best-fit values of $\kappa_{\bar{q}s}$ and $m_{q,s}$, we obtain $f_{K^*}/f_{K}=1.21\,{\rm (SI)}$ and $1.31\,{\rm (SII)}$, being consistent with the experimental value of $1.31$~\cite{PDG,Straub:2015ica}. In addition, we obtain $f_{K^*}^T({\rm 2\,GeV})/f_{K^*}=0.80\,{\rm (SI)}$ and $0.77\,{\rm(SII)}$, which are also comparable with the predictions $f_{K^*}^T({\rm 2\,GeV})/f_{K^*}=0.77$~\cite{Dimopoulos:2008hb}~(LQCD), and $f_{K^*}^T({\rm 2.2\,GeV})/f_{K^*}=0.72$~\cite{Ball:2004rg}, $0.73\pm0.04$~\cite{Ball:2006nr}~(QCDSR). For convenience of comparison, the LQCD and QCDSR, as well as our results for the ratios of decay constants, are summarized in Table~\ref{tab:ratio}.

\begin{table}[t]
\begin{center}
\caption{\label{tab:ratio} \small Summary of our results for the ratios of decay constants. The results given by the LQCD and QCDSR approaches are also listed for comparison. See text for detailed discussion.}
\vspace{0.2cm}
\renewcommand*{\arraystretch}{1.1}
\setlength{\tabcolsep}{6pt}
\begin{tabular}{lccccccccc}
\hline\hline
                   &LQCD             &QCDSR           &SI          &SII\\\hline
$f_{\rho}^T({\rm 2\,GeV})/f_{\rho}$&$0.76$~\cite{Jansen:2009hr}\,, $0.63$~\cite{Braun:2016wnx} &$0.69\pm0.04$~\cite{Ball:2006nr} &$0.78$ &$0.71$\\
$f_{K^*}^T({\rm 2\,GeV})/f_{K^*}$&$0.77$~\cite{Dimopoulos:2008hb} &$0.73\pm0.04$~\cite{Ball:2006nr} &$0.80$ &$0.77$\\
\hline
$f_{D_s}/f_D$      &$1.173\pm0.003$~\cite{PDG}  &$1.170\pm0.023$~\cite{Narison:2015nxh} &$1.129$     &$1.267$\\
$f_{D_s^*}/f_{D^*}$&$1.21\pm0.06$~\cite{Lucha:2014xla}\,, $1.16\pm0.06$~\cite{Becirevic:2012ti}    &$1.16\pm0.04$~\cite{Narison:2015nxh}   &$1.12$     &$1.26$\\
$f_{D^*}/f_{D}$    &$1.078\pm0.036$~\cite{Lubicz:2016bbi,Lubicz:2017asp}  &$1.215\pm0.030$~\cite{Narison:2015nxh} &$1.097$     &$1.131$\\
$f_{D_s^*}/f_{D_s}$&$1.087\pm0.020$~\cite{Lubicz:2016bbi,Lubicz:2017asp}  &$1.19$~\cite{Narison:2015nxh}          &$1.093$     &$1.124$\\
\hline
$f_{B_s}/f_B$      &$1.215\pm0.007$~\cite{PDG}  &$1.154\pm0.021$~\cite{Narison:2015nxh} &$1.166$     &$1.213$\\
$f_{B_s^*}/f_{B^*}$&$1.20$~\cite{Lubicz:2016bbi,Lubicz:2017asp}           &$1.13\pm0.25$~\cite{Narison:2015nxh}   &$1.17$     &$1.21$\\
$f_{B^*}/f_{B}$    &$0.958\pm0.022$~\cite{Lubicz:2016bbi,Lubicz:2017asp}\,, $1.051\pm0.017$~\cite{Becirevic:2014kaa}  &$1.020\pm0.011$~\cite{Narison:2015nxh} &$1.027$     &$1.032$\\
$f_{B_s^*}/f_{B_s}$&$0.974\pm0.010$~\cite{Lubicz:2016bbi,Lubicz:2017asp}          &$0.94$~\cite{Narison:2015nxh}          &$1.028$     &$1.030$\\
\hline\hline
\end{tabular}
\end{center}
\end{table}

\subsection{Heavy-light mesons}

With the best-fit values of $m_{q,s}$ as inputs, we now perform a $\chi^2$-analysis for the holographic parameters in the families of heavy-light mesons, including $D^{(*)}$, $D^{(*)}_s$, $B^{(*)}$ and $B^{(*)}_s$. In this case, because the effect of different $m_{q,s}$ in the two scenarios is trivial, the main difference between SI and SII is now due to whether the heavy-light meson masses are taken into account as constraints or not.

\begin{figure}[t]
\begin{center}
\subfigure[]{\includegraphics[scale=0.5]{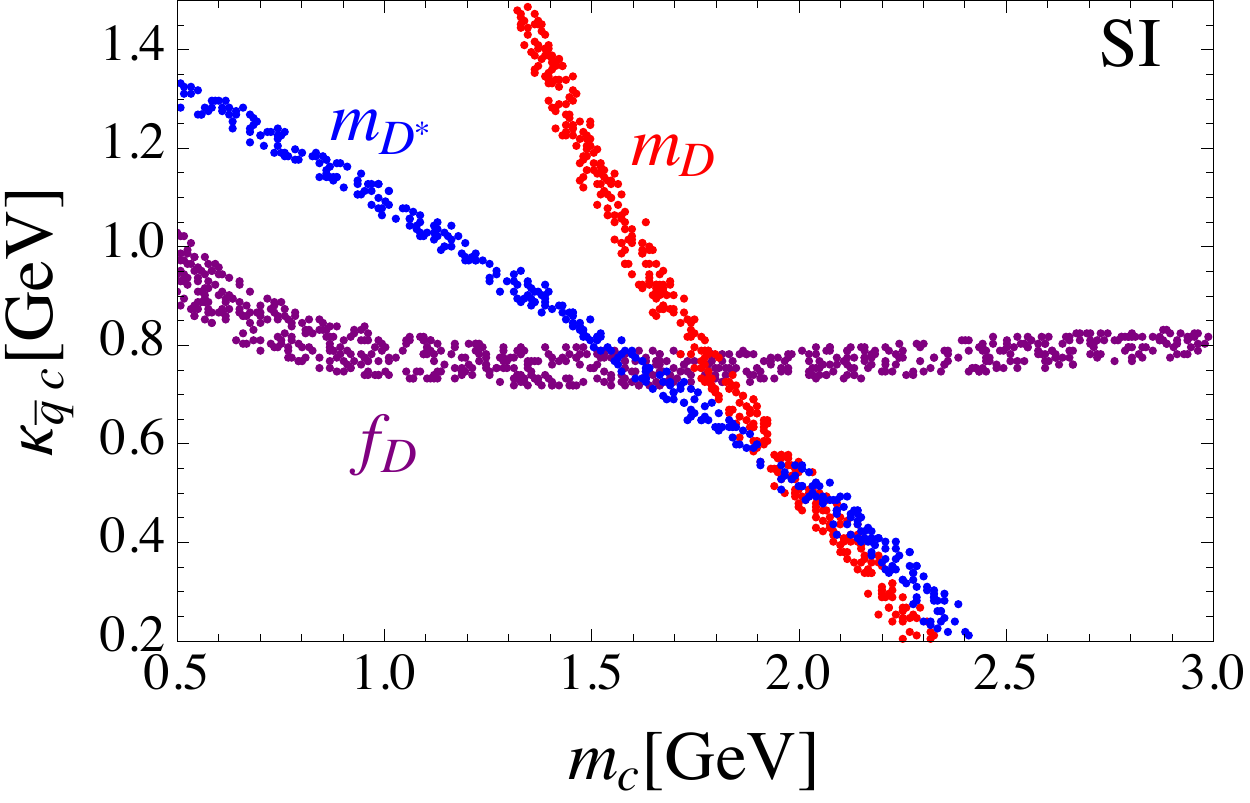}}\qquad
\subfigure[]{\includegraphics[scale=0.5]{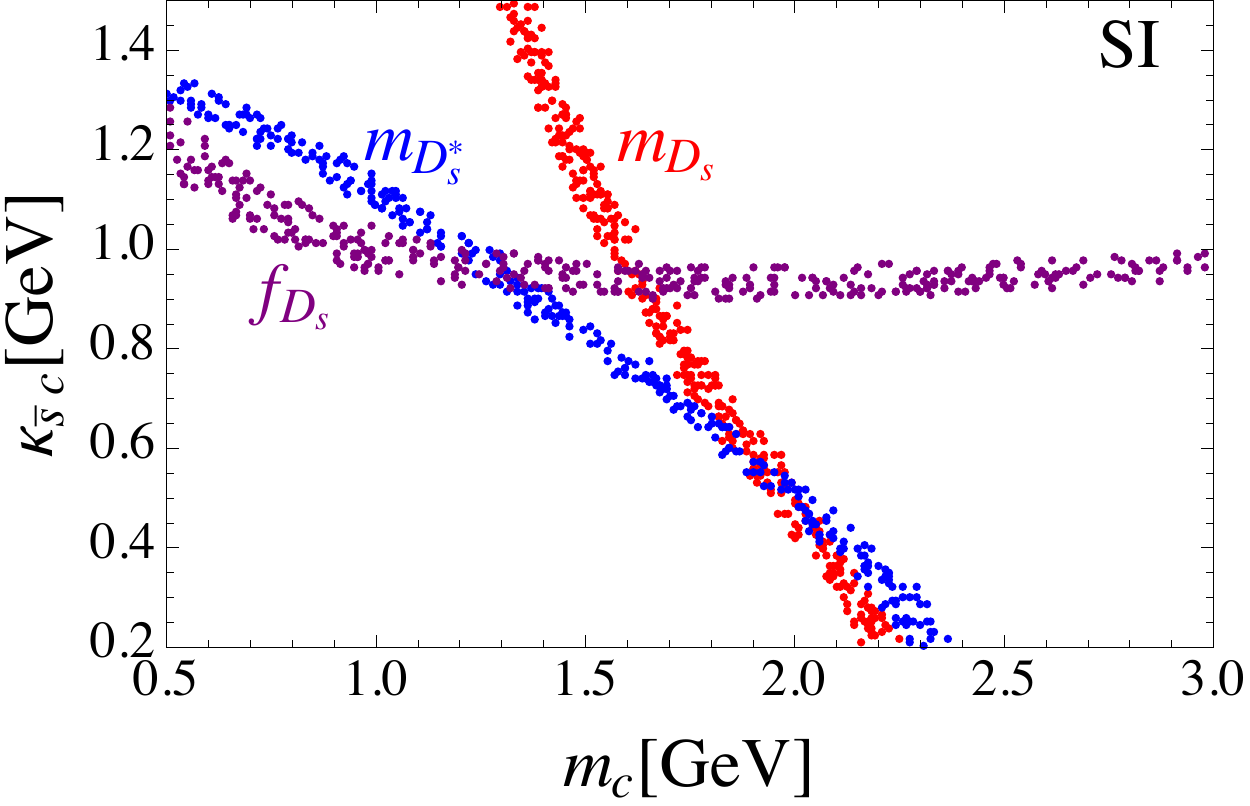}}\\
\subfigure[]{\includegraphics[scale=0.5]{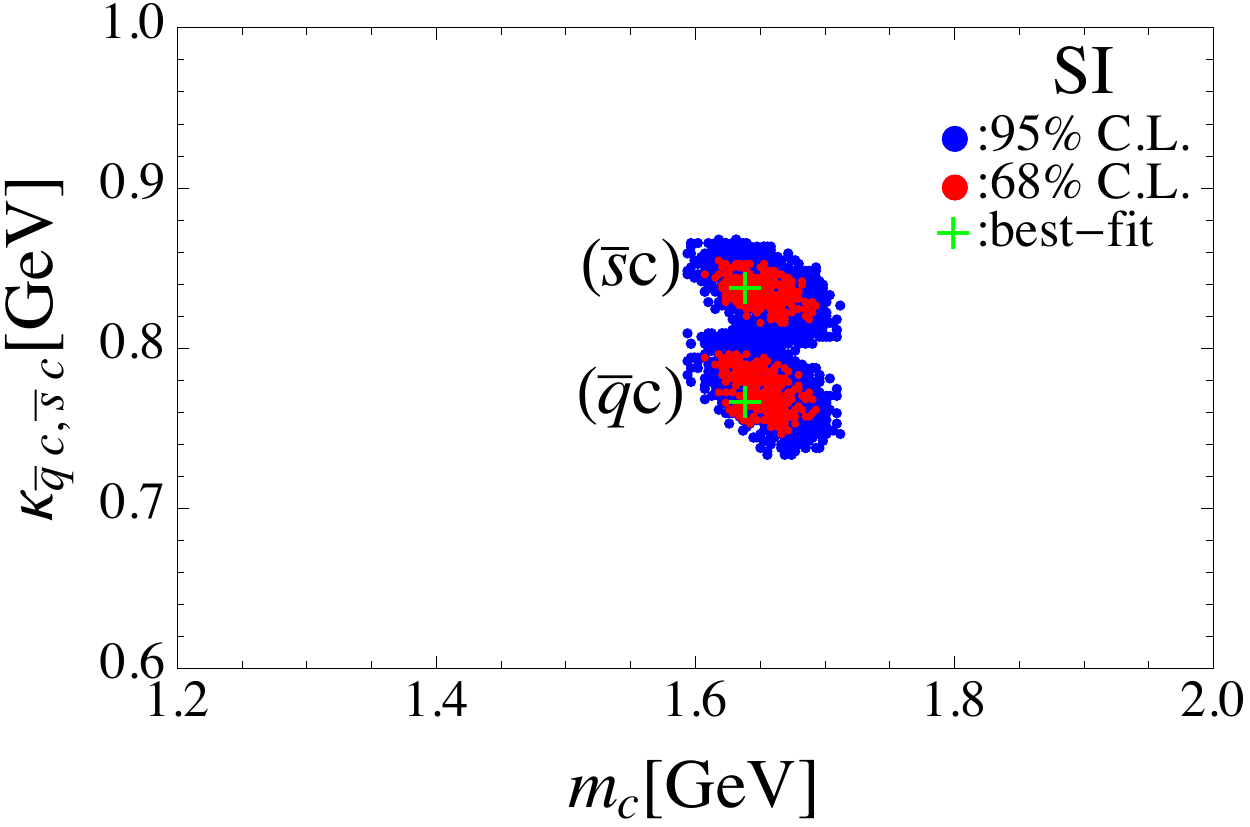}}\qquad
\subfigure[]{\includegraphics[scale=0.5]{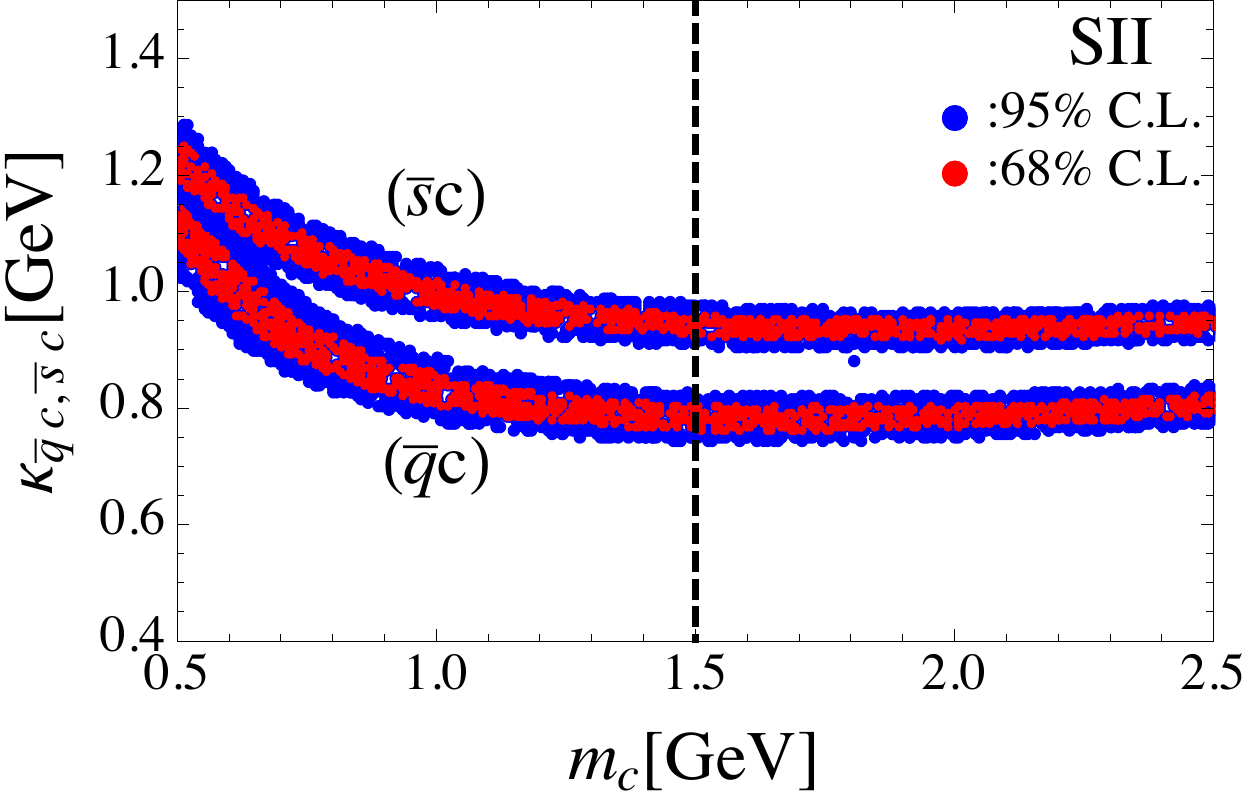}}
\caption{\label{fig:qc} \small The fitted spaces for $\kappa_{\bar{q}c}$, $\kappa_{\bar{s}c}$ and $m_c$ under the separate constraints from the masses and decay constants of $D^{(*)}$ (a) and $D^{(*)}_s$ (b) mesons at $95\%$ C.L. in SI, as well as under their combined constraint in SI (c) and SII (d). The dashed line corresponds to $m_c=1.5~{\rm GeV}$. 
}
\end{center}
\end{figure}

For the $D^{(*)}$ and $D^{(*)}_s$ mesons, the allowed parameter spaces are shown in Fig.~\ref{fig:qc}. From Fig.~\ref{fig:qc}(a), it can be seen that the solution with $\kappa_{\bar{q}c}\sim0.76\,{\rm GeV}$ and $m_c\sim1.6~{\rm GeV}$ is allowed simultaneously by $m_{D}$, $m_{D^*}$ and $f_D$ (there is currently no available data for $f_{D^*}$). Moreover, such a quark mass $m_c\sim1.6\,{\rm GeV}$ is also favored by $m_{D_s}$ and $m_{D^*_s}$ as shown by Fig.~\ref{fig:qc}(b). However, the experimental data on $f_{D_s}$ requires a quite different $\kappa_{\bar{s}c}$. This tension is in fact caused by the observation that the experimental data $f_{D_s}/f_D=1.266$~\cite{PDG} indicates a significant flavour-symmetry breaking effect and hence results in a large difference between $\kappa_{\bar{q}c}$ and $\kappa_{\bar{s}c}$.

Under the combined constraint from the masses and decay constants of the $D^{(*)}$ and $D^{(*)}_s$ mesons, our final fitted results in SI for the holographic parameters $\kappa_{\bar{q}c}$, $\kappa_{\bar{s}c}$ and $m_c$ are shown in Fig.~\ref{fig:qc}(c), with the corresponding numerical results given in Table~\ref{tab:hp}. It is found that the fitted result $\kappa_{\bar{q}c}=0.765^{+0.032}_{-0.018}\,{\rm GeV}$ is in good agreement with that obtained by fitting to the heavy-light hadron spectra~\cite{Dosch:2016zdv}; however $\kappa_{\bar{s}c}=0.836^{+0.020}_{-0.021}\,{\rm GeV}$ is a little bit larger than the result $[0.735, 0.766]\,{\rm GeV}$~\cite{Dosch:2016zdv}, which is due to the effect of $f_{D_s}$ analyzed above. In SII, the fitted results are shown in Fig.~\ref{fig:qc}(d). In this case, because the decay constants $f_{D^*_{(s)}}$ cannot be extracted from experiment for the moment, it is hard for the charm-quark mass to be well bounded; hence we take $m_c=1.5\,{\rm GeV}$ as input, and present in Table~\ref{tab:hp} the fitted results for $\kappa_{\bar{q}c}$ and $\kappa_{\bar{s}c}$.

With the best-fitted holographic parameters as inputs, we further present in Table~\ref{tab:dc} our results for the decay constants of charmed mesons. It can be seen that, except for a slightly smaller $f_{D_s}$ in SI, our results are generally in agreement with those obtained in the LQCD and QCDSR approaches, as well as with the experimental data. In addition, we obtain
\begin{eqnarray}
f_{D_s}/f_{D}=1.129\,{\rm (SI)}\,,1.267\,{\rm (SII)}\,;\qquad f_{D_s^*}/f_{D^*}=1.124\,{\rm (SI)}\,,1.259\,{\rm (SII)}\,,
\end{eqnarray}
which agree with the  LQCD~\cite{PDG,Lucha:2014xla,Becirevic:2012ti} and averaged QCDSR~\cite{Narison:2015nxh} results,
\begin{eqnarray}
&&f_{D_s}/f_{D}=1.173\pm0.003\,{\rm (LQCD)}\,,1.170\pm0.023\,{\rm (QCDSR)}\,;\\
&&f_{D_s^*}/f_{D^*}=1.21\pm0.06,\,1.16\pm0.06\,{\rm  (LQCD)}\,,1.16\pm0.04\,{\rm (QCDSR)}\,.
\end{eqnarray}
Here we should mention that most of the theoretical predictions for $f_{D_s}/f_{D}$ are smaller than the current data, $f_{D_s}/f_{D}\sim1.266$~\cite{PDG}. Finally we obtain
\begin{eqnarray}
f_{D^*}/f_{D}=1.097\,{\rm (SI)}\,,1.131\,{\rm (SII)}\,;\qquad f_{D_s^*}/f_{D_s}=1.093\,{\rm (SI)}\,,1.124\,{\rm (SII)}\,,
\end{eqnarray}
which are also in agreement with the values obtained in the LQCD~\cite{Lubicz:2017asp,Lubicz:2016bbi} and QCDSR~\cite{Narison:2015nxh} approaches
\begin{eqnarray}
&&f_{D^*}/f_{D}=1.078\pm0.036\,{\rm  (LQCD)}\,,1.215\pm0.030\,{\rm (QCDSR)}\,;\\
&& f_{D_s^*}/f_{D_s}=1.087\pm0.020,\,{\rm  (LQCD)}\,,1.19\,{\rm (QCDSR)}\,.
\end{eqnarray}

\begin{figure}[t]
\begin{center}
\subfigure[]{\includegraphics[scale=0.5]{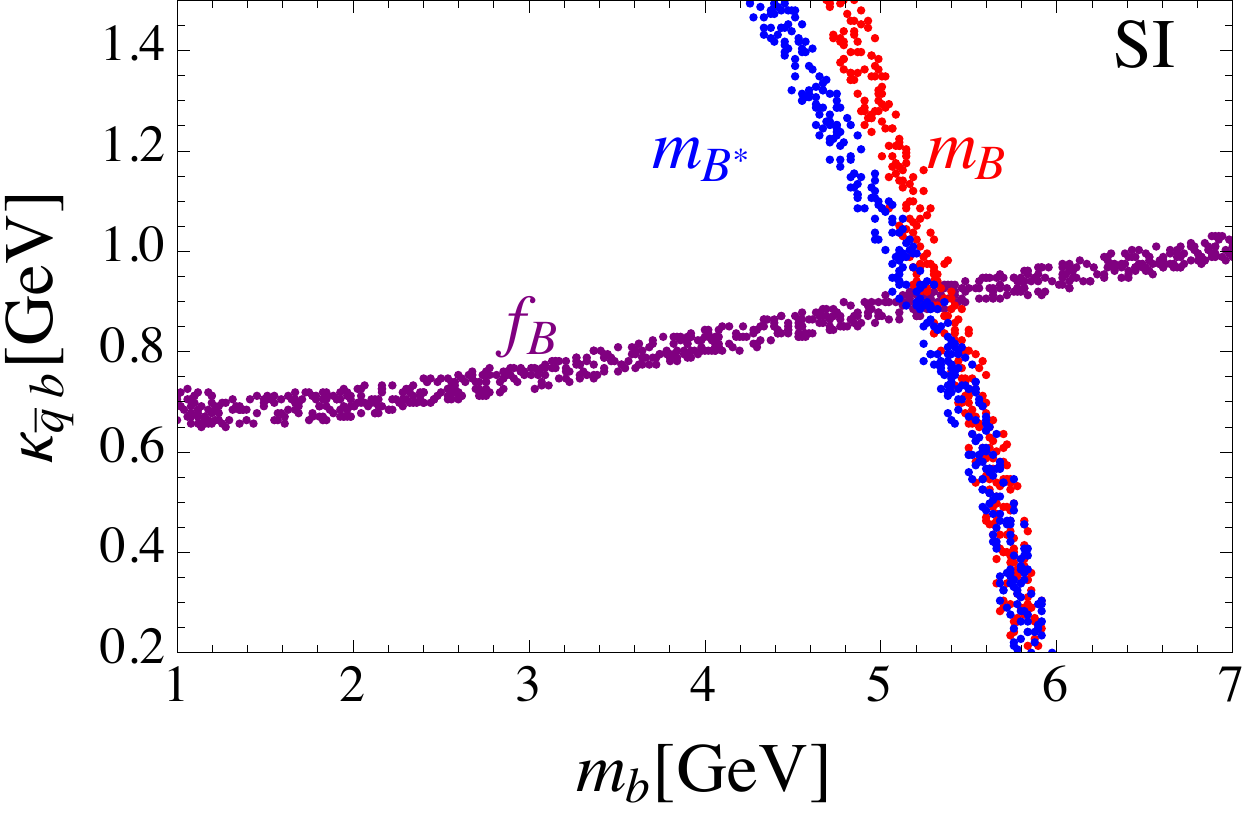}}\qquad
\subfigure[]{\includegraphics[scale=0.5]{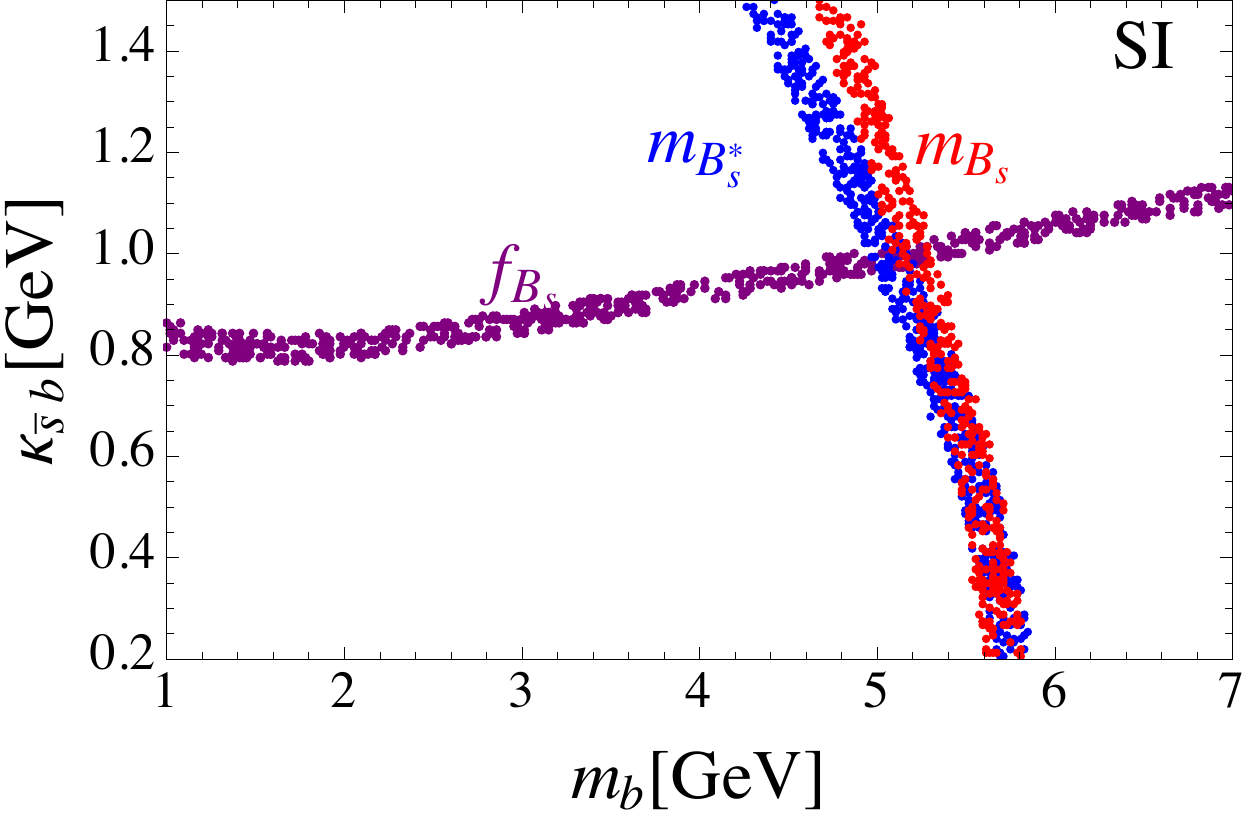}}\\
\subfigure[]{\includegraphics[scale=0.5]{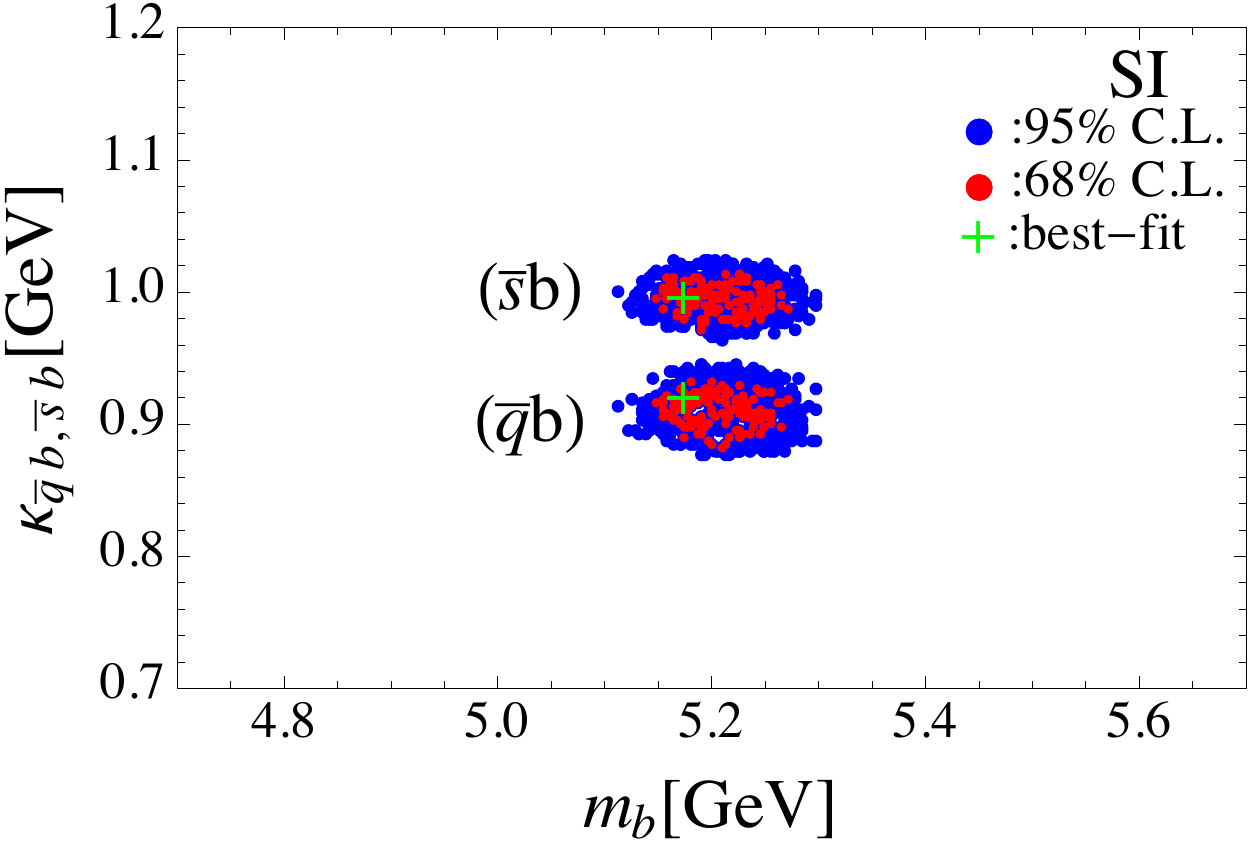}}\qquad
\subfigure[]{\includegraphics[scale=0.5]{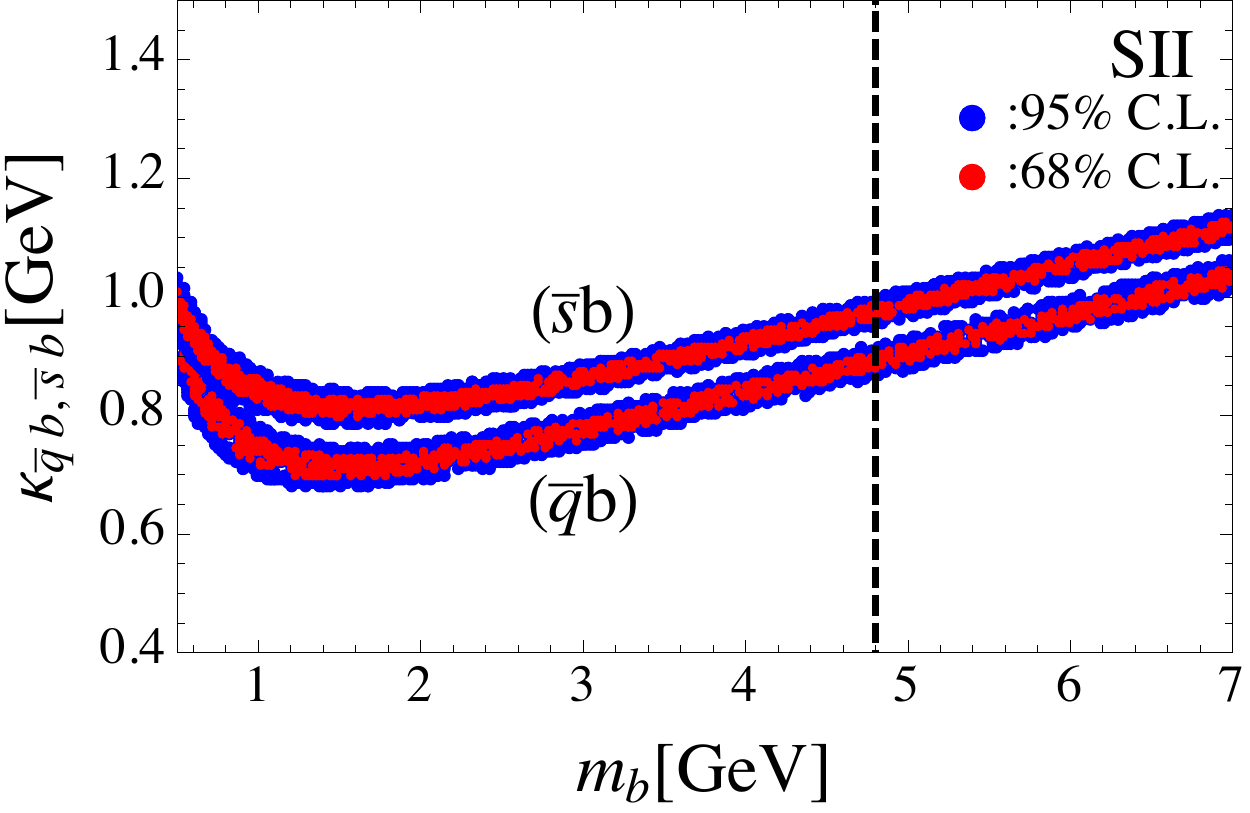}}
\caption{\label{fig:qb} \small The fitted spaces for $\kappa_{\bar{q}b}$, $\kappa_{\bar{s}b}$ and $m_b$ under the separate constraints from the masses and decay constants of $B^{(*)}$ (a) and $B^{(*)}_s$ (b) mesons at $95\%$ C.L. in SI, as well as under their combined constraint in SI (c) and SII (d). The dashed line corresponds to $m_b=4.8~{\rm GeV}$. 
}
\end{center}
\end{figure}

For the $B^{(*)}$ and $B^{(*)}_s$ mesons, the allowed parameter spaces are shown in Fig.~\ref{fig:qb}. It can be seen from Figs.~\ref{fig:qb}(a) and \ref{fig:qb}(b) that the solution $\kappa_{\bar{q}b}\sim0.92\,{\rm GeV}$ and $\kappa_{\bar{s}b}\sim1.00\,{\rm GeV}$ with the same $m_b\sim 5.1$~GeV is allowed simultaneously by $m_{B_{(s)}}$, $m_{B^{*}_{(s)}}$ and $f_{B_{(s)}}$. It is also found that the constraint on $m_b$ is dominated by $m_{B_{(s)}}$ and $m_{B^*_{(s)}}$, while the constraints on $\kappa_{\bar{q}b}$ and $\kappa_{\bar{s}b}$ are dominated by $f_{B}$ and $f_{B_{s}}$, respectively. Different from the case for the charmed mesons, there is no significant tension between the constraints from $B^{(*)}$ and $B^{(*)}_s$ mesons.

Under the combined constraint from the masses and decay constants of $B^{(*)}$ and $B^{(*)}_s$ mesons, our fitting results in SI for the parameters $\kappa_{\bar{q}b}$, $\kappa_{\bar{s}b}$ and $m_b$ are shown in Fig.~\ref{fig:qb}(c), and the corresponding numerical results are given in Table~\ref{tab:hp}, which are also found to be in agreement with those obtained by fitting to the heavy-light hadron spectra~\cite{Dosch:2016zdv}. Similar to the case for the charmed mesons, it is also hard for the holographic parameters in SII to be well bounded due to the lack of data for the decay constants $f_{B^{*}_{(s)}}$, as shown by Fig.~\ref{fig:qb}(d). The fitted results for $\kappa_{\bar{q}b}$ and $\kappa_{\bar{s}b}$ listed in Table~\ref{tab:hp} are, therefore, obtained by taking $m_b=4.8\,{\rm GeV}$.

With the best-fitted holographic parameters as inputs, our results for the decay constants of bottom mesons, which are also listed in Table~\ref{tab:dc}, are generally in agreement with those obtained in the LQCD and QCDSR approaches, but with a few exceptions to be discussed below. In addition, our results for the ratios
\begin{eqnarray}
f_{B_s}/f_{B}=1.166\,{\rm (SI)}\,,1.213\,{\rm (SII)}\,;\qquad f_{B_s^*}/f_{B^*}=1.167\,{\rm (SI)}\,,1.212\,{\rm (SII)}\,,
\end{eqnarray}
agree with the averaged results obtained in the LQCD and QCDSR approaches~\cite{PDG,Lubicz:2016bbi,Lubicz:2017asp,Narison:2015nxh},
\begin{eqnarray}
&&f_{B_s}/f_{B}=1.215\pm0.007\,{\rm  (LQCD)}\,,1.154\pm0.021\,{\rm (QCDSR)}\, ;\\
&&f_{B_s^*}/f_{B^*}=1.20\,{\rm  (LQCD)}\,,1.13 \pm0.25\,{\rm (QCDSR)}\,.
\end{eqnarray}
On the other hand, we obtain
\begin{eqnarray}
f_{B^*}/f_{B}=1.027\,{\rm (SI)}\,,1.032\,{\rm (SII)}\,;\qquad f_{B_s^*}/f_{B_s}=1.028\,{\rm (SI)}\,,1.030\,{\rm (SII)}\,,
\end{eqnarray}
which are approximately equal to but a little bit larger than one. Comparing with the results obtained in the LQCD and QCDSR approaches~\cite{Lubicz:2016bbi,Lubicz:2017asp,Narison:2015nxh,Becirevic:2014kaa},
\begin{eqnarray}
&&f_{B^*}/f_{B}=0.958\pm0.022\,,1.051\pm0.017{\rm  (LQCD)}\,,1.020\pm 0.011\,{\rm (QCDSR)}\,;\\
&& f_{B_s^*}/f_{B_s}=0.974\pm0.010\,{\rm (LQCD)}\,,0.94\,{\rm (QCDSR)}\,,
\end{eqnarray}
we can find the following differences: In our approach, for a given ($q_1\bar{q}_2$) state, the relation $f_V/f_P>1$ is always satisfied and can reach to one only in the heavy quark limit, which has been analyzed in the last section and can also be seen from Eqs.~\eqref{eq:redcp} and \eqref{eq:redcv}. While the QCDSR predictions~\cite{Narison:2015nxh} support $f_{B^*}/f_{B}>1$, the relation $f_{B^*}/f_{B}<1$ is predicted by most of the LQCD evaluations, for instance, in Refs.~\cite{Lubicz:2016bbi,Colquhoun:2015oha,Lubicz:2017asp}, but with the exception of that obtained with $N_f=2$ dynamical quarks~\cite{Becirevic:2014kaa}. In addition, both the LQCD and QCDSR approaches support the relation $f_{B_s^*}/f_{B_s}<1$. As a consequence, more precise information from both theoretical and experimental sides is needed to resolve these discrepancies.

\section{Conclusion}

In this paper, the decay constants of light and heavy-light pseudoscalar and vector mesons have been evaluated with the improved soft-wall holographic wavefunctions, which are now modified to take the effects of both quark masses and dynamical spins into account properly. Taking the masses and measured decay constants of these mesons as constraints, we have performed detailed $\chi^2$-analyses to determine the holographic parameters, the mass-scale parameter $\kappa$ and the quark masses, in two different scenarios. With the best-fitted parameters as inputs, we also presented our theoretical results for the decay constants as well as some important ratios among them. Our main findings can be summarized as follows:
\begin{itemize}
\item Our results for the decay constants in the holographic QCD formalism, especially for the ratio $f_V/f_P$, can be significantly improved once the dynamical spin effects are taken into account by introducing a helicity-dependent wavefunction.

\item Our fitted results in SI for the mass-scale parameters, $\kappa_{\bar{q_2}q_1}$, as summarized in Table~\ref{tab:hp}, are generally in agreement with those obtained by fitting to the Regge trajectories~\cite{Brodsky:2014yha,Dosch:2016zdv}. With the determined holographic parameters as inputs, our results for the decay constants agree well with the data, but with some tensions for $f_{\rho}$ and $f_{D_s}$.

\item Most of our theoretical results are also in agreement with those obtained in the LQCD and QCDSR approaches. The only observed tension between these methods and ours lies in the ratios $f_{B_{(s)}^*}/f_{B_{(s)}}$, which are predicted to be smaller than one in LQCD, but a little bit larger than one in this work and can reach to one only in the heavy quark limit.
\end{itemize}

\section*{Acknowledgements}
This work is supported by the National Natural Science Foundation of China (Grant Nos. 11475055, 11675061 and 11435003). Q.~Chang is also supported by the Foundation for the Author of National Excellent Doctoral Dissertation of P.~R.~China (Grant No. 201317), the Program for Science and Technology Innovation Talents in Universities of Henan Province (Grant No. 14HASTIT036), the Excellent Youth Foundation of HNNU. X.~Li is also supported in part by the self-determined research funds of CCNU from the colleges' basic research and operation of MOE~(CCNU18TS029).


\clearpage

\end{CJK*}
\end{document}